\documentclass{aa}  

\usepackage{graphicx}
\usepackage{xcolor}
\usepackage{txfonts}
\begin{document} 

   \title{
   Physical properties and trigonometric distance of the peculiar dwarf WISE\,J181005.5$-$101002.3}

%   \subtitle{Full characterisation of WISE1810}
   \titlerunning{
   Physical properties and distance of the peculiar dwarf WISE J181005.5$-$101002.3}

   \author{N.\ Lodieu \inst{1,2}
       \and
       M. R.\ Zapatero Osorio \inst{3}
       \and
       E.\ L.\ Mart\'in \inst{1,2,4}
        \and
       R. Rebolo L\'opez \inst{1,2,4}
        \and
       B.\ Gauza \inst{5,6} %ORCID 0000-0001-5452-2056
        }

   \institute{Instituto de Astrof\'isica de Canarias (IAC), Calle V\'ia L\'actea s/n, E-38200 La Laguna, Tenerife, Spain \\
       \email{nlodieu@iac.es}
       \and
       Departamento de Astrof\'isica, Universidad de La Laguna (ULL), E-38206 La Laguna, Tenerife, Spain
       \and
       Centro de Astrobiolog\'ia (CSIC-INTA), Ctra. Ajalvir km 4, 28850, Torrej\'on de Ardoz, Madrid, Spain
       \and
       Consejo Superior de Investigaciones Cient\'ificas (CSIC), Madrid, Spain
       \and
        Centre for Astrophysics Research, School of Physics, Astronomy and Mathematics, University of Hertfordshire, College Lane, Hatfield AL10 9AB, UK
        \and
        Janusz Gil Institute of Astronomy, University of Zielona G\'ora, Lubuska 2, 65–265 Zielona G\'ora, Poland
       %\and
       }

   \date{\today{},\today{}}

% \abstract{}{}{}{}{} 
% 5 {} token are mandatory
 
  \abstract
  % context heading (optional)
  % {} leave it empty if necessary  
  {}
  % aims heading (mandatory)
   {Our goal is to characterise the physical properties of the metal-poor brown dwarf population. In particular, we focus on the recently discovered peculiar dwarf WISE1810055$-$1010023.  
   }
  % methods heading (mandatory)
   {We collected optical $iz$ and near-infrared $J$-band imaging on multiple occasions over 1.5 years to derive accurate trigonometric parallax and proper motion of the metal-depleted ultra-cool dwarf candidate WISE\,J1810055$-$1010023. We also acquired low-resolution optical spectroscopy (0.6$-$1.0 $\mu$m) and new infrared (0.9$-$1.3 $\mu$m) spectra of WISE\,J1810055$-$1010023 that were combined with our photometry, other existing data from the literature and our trigonometric distance to determine the object's luminosity from the integration of the observed spectral energy distribution covering from  0.6 through 16 $\mu$m. We compared the full optical and infrared spectrum with state-of-the-art atmosphere models to further constrain its effective temperature, surface gravity and metallicity.
   }
  % results heading (mandatory)
   {WISE\,J1810055$-$1010023 is detected in the $iz$ bands with AB magnitudes of $i$\,=\,23.871$\pm$0.104 and $z$\,=\,20.147$\pm$0.083 mag in the Panoramic Survey Telescope and Rapid Response System (PanSTARRS) system. It does not show any obvious photometric variability beyond 0.1$-$0.2 mag in any of the $z$- and $J$-band filters. The very red $z-J \approx 2.9$ mag colour is compatible with an ultra-cool dwarf nature. Fitting for parallax and proper motion, we measure a trigonometric parallax of 112.5\,$^{+8.1}_{-8.0}$ mas for WISE\,J1810055$-$1010023, placing the object at only  8.9$^{+0.7}_{-0.6}$ pc, about three times closer than previously thought. We employed Monte Carlo methods to estimate the error on the parallax and proper motion. The object's luminosity was determined at log\,$L/L_\odot$\,=\,$-5.78 \pm 0.11$ dex. From the comparison to atmospheric models, we infer a likely metallicity of [Fe/H] $\approx -1.5$ and an effective temperature cooler than 1000 K. 
   The estimated luminosity and temperature of this object are below the known substellar limit. Despite its apparent low metallicity, we derive space motions that are more typical of the old disc than the halo of the Milky Way. We confirm that WISE\,J1810055$-$1010023 has an ultra-cool  temperature and belongs to a new class of objects with no known spectral counterparts among field L- and T-type dwarfs. 
   }
  % conclusions heading (optional), leave it empty if necessary 
   {WISE\,J1810055$-$1010023 is a very special substellar object and represents a new addition to the 10 pc sample. The optical to near-infrared spectra show strong features due to water vapour and H$_2$ collision induced absorption. Our trigonometric distance has strong implications on the density of metal-poor brown dwarfs in the solar vicinity, which may be higher than that of metal-poor stars.
    }


   \maketitle
%
%

%%%%%%%%%%%%%%%%%%%%%%%%%%%%%%%%%%%%%%%%%%%%
%%%%% Introduction %%%%%
%%%%%%%%%%%%%%%%%%%%%%%%%%%%%%%%%%%%%%%%%%%%
%
\section{Introduction}
\label{esdT:intro}

The physical properties of substellar objects evolve with time. Hence, finding brown dwarfs of
different ages and metallicities is critical to tracing their evolutionary paths and understanding the processes of substellar formation. Since the 
discoveries of the first two unambiguous brown dwarfs announced in 1995 \citep{rebolo95,nakajima95}, the field has grown significantly with thousands of substellar objects with a wide range of physical properties: mass, age, atmospheric composition \citep[see review by][]{kirkpatrick05}. Nonetheless, metal-poor dwarfs at or below the hydrogen-burning limit still remain scarce. 

Metal-poor, low-mass stars with spectral types later than M7 are important tracers of the Galactic structure and chemical enrichment due to their long nuclear burning lifetimes. Metal-depleted brown dwarfs, which lack important nuclear burning and can be even cooler than the smallest stars, are also key \citep{lodieu18c}. All of them, known as ultra-cool subdwarfs \citep{kirkpatrick97}, enable the study of early star formation channels, nucleo-synthesis pathways, and Galactic assembly because they 
represent unique reservoirs of a pristine material that is somewhat processed, but with a far lower metallicity than the Sun, and produced by the Big Bang.
They are also relevant objects to constrain atmospheric models at low temperatures and deficient chemical compositions
\citep{rajpurohit14,rajpurohit16a,zhang17a,zhang18b,lodieu19a,schneider20a,meisner20b,meisner21,brooks22a}. 
Subdwarfs tend to exhibit thick disc or halo kinematics, high proper motions, and
high heliocentric velocities \citep{gizis97a}, although it is not necessarily the case for all metal-poor ultra-cool dwarfs \citep[e.g.\ Figure 17 in][]{zhang17a}.  
Metal-depleted low-mass stars, despite the higher temperatures caused by the dearth of metals in their atmospheres, are less luminous than their solar-metallicity
counterparts due to  their smaller radii \citep{baraffe97,allard12,gonzales21,gerasimov22}. As for low-metallicity brown dwarfs below the hydrogen burning-mass limit, there is no recent theoretical work in the literature we can rely on for the predictions of their sizes and temperatures, but an approximate observational stellar-substellar boundary is drawn in Figure 1 of \citet{zhang17b}.
From current spectroscopic observations of low-metallicity M- and L-type dwarfs at optical and near-infrared wavelengths, the spectra show stronger metal-hydride absorption and weaker
metal lines than solar-metallicity dwarfs of similar classification, as well as blue infrared colours caused by collision-induced H$_{\rm 2}$ absorption
\citep{lodieu17a}. The adopted classification for metal-poor dwarfs \citep{kirkpatrick14,zhang17a} is divided into three categories \citep{lepine07c} with well-established iron abundances \citep{zhang17a,lodieu19a}: subdwarfs 
(sd; [Fe/H]\,=\,$-$ 0.5$\pm$0.5 dex), extreme subdwarfs (esd; [Fe/H]\,=\,$-$1.5$\pm$0.5 dex), and ultra-subdwarfs 
(usd; [Fe/H]\,=\,$-$2.0$\pm$0.5 dex). 

Only a few metal-poor brown dwarf candidates have been announced so far in isolation
\citep{burgasser03b,murray11,burningham14,greco19a,meisner20a,meisner20b,meisner21,brooks22a} or as companions \citep{burningham10a,scholz10a,pinfield12,mace13} but all have metallicities likely higher 
than [Fe/H] = $-$0.7 dex. The first extremely metal-poor dwarf candidates ([Fe/H]\,$\leq$\,$-$1.0 dex) located in the substellar transition zone have been recently identified in the Wide-Field Infrared Survey Explorer
\citep[WISE;][]{wright10} catalogue by \citet{schneider20a}. Their near-infrared spectra cannot be reproduced by any current models or observed field T dwarf templates \citep{burgasser06a}. Their photometry appears equally peculiar with distinctive blue $W1-W2$ colours \citep{schneider20a,meisner20b,goodman21a}. \citet{kirkpatrick21b} reported an even fainter enigmatic metal-poor candidate potentially classified as a Y-type dwarf \citep{cushing11,kirkpatrick12} and nicknamed ``the Accident''.

\citet{schneider20a} announced WISE1810055$-$1010023 (hereafter WISE1810) as one of the first 
extremely metal-poor ultra-cool subdwarfs confirmed as a brown dwarf based on its infrared spectrum. WISE1810 is relatively 
bright at infrared wavelength ($J$\,=\,17.26 mag, $W2$\,=\,12.48 mag) with a near-infrared spectrum
best fit by an observational T0 template and with an effective temperature of 1300$\pm$100\,K, 
$\log$(g)\,=\,5.0 dex, and metallicity below $-$1.0 dex based on model fit. According to \citet{schneider20a}, WISE1810 appears to be located
at a spectrophotometric distance of 14$-$67 pc  taking into account the magnitude--distance relations of field, most likely solar-metallicity, T dwarfs.
These authors also inferred a mass range of 0.075$-$0.080 M$_{\odot}$ from known 
L and T subdwarfs with published effective temperatures and metallicities \citep{kirkpatrick14,zhang17a}.

This paper discusses the properties of WISE1810, one of the first two extremely metal-poor brown dwarf candidates announced by the WISE team \citep{schneider20a}. Section \ref{esdT:Obs_phot} presents optical (300--1000 nm) and infrared $J$-band photometry of WISE1810 obtained with a suite of instruments. Section \ref{esdT:Obs_spec} reports on a low-resolution optical (0.7--1.0 micron) spectrum of WISE1810 and an improved 900$-$1330 nm infrared spectrum. Section \ref{esdT:Obs_parallax} introduces the first ground-based trigonometric parallax for WISE1810 by combining archival data with new imaging data presented in this manuscript.
Section \ref{esdT:properties} derives the physical properties of WISE1810 using all observational information available to us and discusses the implications of WISE1810 with regard to the density of metal-poor brown dwarfs in the solar neighbourhood.
 
%
%%%%%%%%%%%%%%%%%%%%%%%%%%%%%%%%%%%%%
%%%%% Table: Logs for WISE1810 %%%%%
%%%%%%%%%%%%%%%%%%%%%%%%%%%%%%%%%%%%%
%
\begin{table}
 \centering
 \caption[]{Summary of new photometric and spectroscopic observations of WISE1810.
 }
\tiny
 \begin{tabular}{@{\hspace{-0.1mm}}l @{\hspace{2mm}}l @{\hspace{2mm}}c @{\hspace{1mm}}c @{\hspace{1mm}}c @{\hspace{1mm}}c@{\hspace{0mm}}}
 \hline
 \hline
Instr &  Date   &   UT  &  Exp$.$ time  & Seeing & Fil/Grating  \cr
      &  & (hh:mm) &  (s)   &    ($"$) &       \cr
 \hline
ALFOSC    & 2020 Aug 18 & 21:55 & 10$\times$300 & 0.6 & $z$ \cr
ALFOSC    & 2021 Feb 16 & 05:35 & 25$\times$150 & 1.1 & $z$ \cr
OSIRIS    & 2020 Sep 06 & 21:34 & 1$\times$45   & 0.9 & $z$ \cr
OSIRIS    & 2020 Sep 06 & 21:50 & 2$\times$1800   & 0.9 & R1000R \cr
OSIRIS    & 2020 Sep 19 & 21:01 & 5$\times$50   & 0.7 & $z$ \cr
OSIRIS    & 2020 Sep 19 & 21:27 & 2$\times$1800   & 0.7 & R1000R \cr
OSIRIS    & 2020 Sep 28 & 20:59 & 5$\times$50   & 0.9 & $z$ \cr
OSIRIS    & 2020 Oct 09 & 20:21 & 10$\times$50  & 0.9 & $z$ \cr
OSIRIS    & 2020 Oct 23 &       &               &.     & $z$ \cr
OSIRIS    & 2020 Oct 25 & 19:59 & 10$\times$30  & 1.0 & $z$ \cr
OSIRIS    & 2020 Nov 08 & 19:38 & 10$\times$25  & 1.0 & $z$ \cr
Omega2000 & 2021 Mar 25 & 04:20 & 9$\times$30  & 1.2 & $J$ \cr
Omega2000 & 2021 Apr 21 & 03:05 & 9$\times$30  & 1.3 & $J$ \cr
Omega2000 & 2021 Jul 01 & 02:46 & 9$\times$30  & 1.3 & $J$ \cr
Omega2000 & 2021 Jul 28 & 23:58 & 9$\times$30  & 1.1 & $J$ \cr
Omega2000 & 2021 Aug 24 & 21:25 & 9$\times$30  & 1.6 & $J$ \cr
Omega2000 & 2021 Oct 19 & 18:58 & 9$\times$30  & 1.0 & $J$ \cr
EMIR      & 2021 Mar 04 & 06:35 & 7$\times$10  & 0.9 & $J$ \cr
EMIR      & 2021 Apr 22 & 04:32 & 7$\times$10  & 0.9 & $J$ \cr % BAD
EMIR      & 2021 May 26 & 01:36 & 7$\times$10  & 0.9 & $J$ \cr
EMIR      & 2021 May 29 & 04:45 & 7$\times$10  & 0.9 & $J$ \cr
EMIR      & 2021 Jun 22 & 01:48 & 7$\times$10  & 0.9 & $J$ \cr
EMIR      & 2021 Jun 26 & 01:15 & 20$\times$360  & 0.9 & $YJ$ \cr
EMIR      & 2021 Jun 26 & 23:25 & 1$\times$10  & 0.8 & $J$ \cr
EMIR      & 2021 Jun 26 & 23:43 & 20$\times$360  & 0.8 & $YJ$ \cr
EMIR      & 2021 Jul 18 & 23:10 & 7$\times$10  & 0.6 & $J$ \cr
EMIR      & 2021 Aug 07 & 21:21 & 7$\times$10  & 1.0 & $J$ \cr
EMIR      & 2021 Aug 25 & 22:33 & 7$\times$10  & 0.8 & $J$ \cr
HiPERCAM  & 2021 May 05 & 03:30 & 2$\times$12.87$+$6$\times$32.97    & 0.6 & $u$  \cr
HiPERCAM  & 2021 May 05 & 03:30 & 2$\times$12.87$+$6$\times$32.97    & 0.6 & $g$  \cr
HiPERCAM  & 2021 May 05 & 03:30 & 2$\times$12.87$+$6$\times$32.97    & 0.6 & $r$  \cr
HiPERCAM  & 2021 May 05 & 03:30 & 27$\times$12.87$+$60$\times$32.87  & 0.6 & $i$ \cr
HiPERCAM  & 2021 May 05 & 03:30 & 27$\times$12.87$+$60$\times$32.87  & 0.6 & $z$ \cr
\hline
 \label{tab_esdT:tab_logs}
 \end{tabular}
\end{table}
%

%
%%%%%%%%%%%%%%%%%%%%%%%%%%%%%%%%%%%%%%%%%%%%
%%%%% Observations: photometry %%%%%
%%%%%%%%%%%%%%%%%%%%%%%%%%%%%%%%%%%%%%%%%%%%
%
\section{Observations}
\label{esdT:Obs_phot}

 We observed WISE1810 photometrically and spectroscopically on multiple occasions with the purpose of determining its trigonometric parallax, optical to infrared spectral energy distribution, and photometric variability. The observing logs of all photometric and spectroscopic campaigns are given in Table~\ref{tab_esdT:tab_logs}. 
 The observations and data reduction steps are properly described in this section. The derivation of the observed photometry and relative astrometry is explained in Sections~\ref{phot_analysis} and~\ref{esdT:Obs_parallax}.

\subsection{Optical and near-infrared imaging}

\subsubsection{NOT ALFOSC}
\label{esdT:Obs_imaging_NOT}

We secured optical images of WISE1810 with the Alhambra Faint Object Spectrograph and Camera (ALFOSC) 
on the 2.5m Nordic Optical Telescope (NOT) during a service night on 18 August 2020 between UT\,=\,20h54m and 21h55m under programme number 61\,299 (PI Lodieu). ALFOSC has a field of view of 6.4$\times$6.4 arcmin$^2$ in imaging mode and a nominal pixel size of 0\farcs2138. We used the $z$-band filter $z'\_SDSS\,832\_LP$ with a central wavelength of 840 nm.
The sky was clear and dark with a seeing of 0\farcs5--0\farcs6. 
We collected a total of ten individual exposures of 360 s each in staring mode.

\begin{figure}
   \centering
   \includegraphics[width=\linewidth, angle=0]{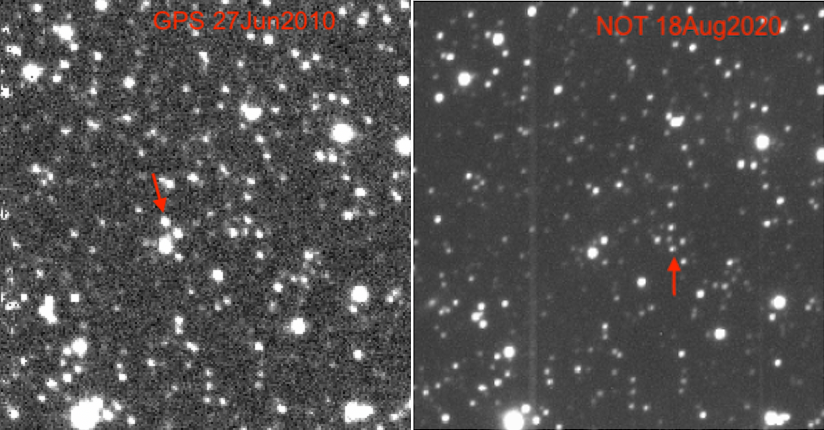}
   \caption{
   Images of WISE1810 from the UKIDSS Galactic Plane Survey (left; 27 June 2010) and 
   NOT ALFOSC (right; 18 August 2020) in $J$ and $z$ filters, respectively. 
   The object is indicated by a red arrow.
   North is up and east left. Each image is about 80 arcsec across.
   }
   \label{fig_esdT:@ISE1810images}
\end{figure}

We carried out the data reduction under the IRAF\footnote{IRAF is distributed by the National Optical Astronomy
Observatories, which are operated by the Association of Universities for Research in
Astronomy, Inc., under cooperative agreement with the National Science Foundation} environment
\citep{tody86,tody93}. We subtracted the median-combined bias frame from each flat-field image and from each 
individual science frame. Then, we  divided each science frame by a normalised median-combined 
flat field. Afterwards, we averaged all science images removing the lowest and highest values (Fig.\ \ref{fig_esdT:@ISE1810images}). 

\subsubsection{GTC OSIRIS }
\label{esdT:Obs_imaging_GTC_OSIRISz}

We obtained Sloan $z$-band photometry of WISE1810$-$10 with the OSIRIS
Optical System for Imaging and low-intermediate Resolution Integrated Spectroscopy \citep[OSIRIS;][]{cepa00}
instrument mounted on the 10.4m Gran Telescopio de Canarias (GTC).
OSIRIS is equipped with two 2048$\times$4096 Marconi CCD42-82 detectors working at optical wavelengths 
and offering a field of view of approximately 7$\times$7 arcmin$^2$ with an unbinned, nominal pixel scale of 0\farcs125\@.

We collected five epochs as part of Director Discretionary Time allocation (DDT) programme GTC2020-163 (PI Lodieu) on 19 and 28 September 2020, 9 and 25 October 2020, and 8 November 2020 in service mode. All data were acquired with the $z$-band filter centred at 0.895 $\mu$m. The sky was clear and the seeing measured on the images was subarcsec in all datasets, as requested during the phase 2 preparation of the observing blocks. We set the on-source integration to 50 s with a five-point dither pattern of 3\farcs5 to avoid overlap with nearby sources as much as possible due to the high density of this field close to the Galactic plane. On the last observing epoch we collected ten dithered images of 25 s each, due to the very high air mass of WISE1810\@. To increase the number of astrometric measurements, we also incorporated the acquisition images of the spectroscopic observing block obtained with OSIRIS on 2020 September 6 as part of DDT programme GTC2020-161 (PI Lodieu; see Section \ref{esdT:Obs_spec}).

We reduced all data with IRAF\@.
We subtracted the bias from each individual frame and divided by a normalised flat field.
We median-combined all frames correcting for the offsets set by the dithering pattern. 

%
%%%%%%%%%%%%%%%%%%%%%%%%%%%%%%%%%%%%%%%%%%%%
%%%%% Figure %%%%%
%%%%%%%%%%%%%%%%%%%%%%%%%%%%%%%%%%%%%%%%%%%%
%
\subsubsection{GTC HiPERCAM}
\label{esdT:Obs_imaging_GTC_HiPERCAM}
We collected High PERformance CAMera \citep[HiPERCAM;][]{dhillon21a} data in service mode as part of programme GTCMULTIPLE3A-21A (PI Lodieu) on the night of 9 May 2021\@. The same observing block was repeated because there was some spare time. The conditions were photometric with a clear sky, a seeing of 0\farcs6, and dark time. In total, we collected eight images in $ugr$ and 87 in $iz$ Sloan filters, totalling 223.5\,s and 2326\,s on-source exposure time, respectively.

HiPERCAM has been a visitor instrument at the GTC since February 2018\@. This instrument uses 
four beam-splitters to image simultaneously the same field in five optical filters
($ugriz$) covering the 300--1000 nm wavelength range in one single shot. Each detector has four read-out
quadrants of 1024$\times$512 pixels with a nominal size of 0\farcs081 covering a field of view of 2.8\,$\times$\,1.4 arcmin$^2$.
The main advantage of HiPERCAM in our study is the availability of simultaneous $ugriz$ photometry with good sensitivity in the blue. 
The automatic HiPERCAM pipeline was run on a dedicated machine and we obtained
images in each individual filter by running the {\tt{joinup}} script 
in the Python environment\footnote{\url{http://deneb.astro.warwick.ac.uk/phsaap/hipercam/docs/html/}}. 

\subsubsection{Calar Alto Omega2000}
\label{esdT:Obs_imaging_Omega2000}

We collected new $J$-band photometry with Omega2000 on the Calar Alto
3.5m telescope over two semesters (programmes H20-3.5-020 and F21-3.5-010; PI Lodieu) 
to further constrain the trigonometric parallax of WISE1810.
Omega2000 employs a HAWAII-2 2048$\times$2048 pixel detector sensitive
in the 1000--2500 nm wavelength range with a nominal pixel size of 0\farcs45,
yielding a field of view of 15.4 $\times$ 15.4 arcmin$^2$ \citep{kovacs04}.

We secured six observing epochs on 2021 March 25, April 21, July 01, July 28, August 24, and October 19 (Table \ref{tab_esdT:tab_logs}).
The seeing measured on the final images was generally between 1\farcs0 and 1\farcs6. The 2021 August 24 data have the worst seeing.
We set the on-source integration time to 10\, s with three integrations each followed 
by a nine-point dither pattern, yielding a total on-source exposure time of 4.5 min per observing epoch.

We reduced the data in a standard manner under the IRAF environment, following steps that were very similar to those used for the other near-infrared instruments.
We created a median-combined flat-field image subtracting the dome ON and OFF
images taken during the afternoon preceding the night or morning at the end of
the night. We also median-combined eight of the nine dithered images that we subtracted from the unused image to remove the sky contribution. Then we
combined with an average method the nine dithered frames applying the offsets
calculated from a bright unsaturated star in a central part of the detector.
None of the Omega2000 images was specifically calibrated astrometrically as it is not
required for our astrometric analysis (see Section \ref{esdT:Obs_parallax}).

\subsubsection{GTC EMIR }
\label{esdT:Obs_imaging_GTC_EMIR_J}

We also obtained complementary $J$-band photometry with the Espectrografo Multiobjeto Infra-Rojo \citep[EMIR;][]{garzon07}
mounted on the Naysmith-A focus of the GTC as part of programme GTCMULTIPLE3A-21A (PI Lodieu). 
EMIR is equipped with a 2048$\times$2048 pixel Teledyne HAWAII-2 HgCdTe detector offering a 
6.67$\times$6.67 arcmin$^{2}$ field of view with a nominal pixel scale of 0\farcs2 on the sky. 
All observations were carried out in service mode under clear skies and seeing better than 1\farcs0 to avoid as much as possible any contamination from nearby stars. We collected a total of nine epochs between 2021 March and 2021 August (Table \ref{tab_esdT:tab_logs}).

We reduced the EMIR images with the official EMIR pipeline PyEMIR\footnote{\url{https://pyemir.readthedocs.io/}}
to produce a final stacked image at each epoch without any astrometric calibration. 
The pipeline removes the flat-field contribution. 
We calculated the dithering offsets within the IRAF environment, which we input into the pipeline as one of the 
options to apply the offsets between the standard seven-point dither pattern in the EMIR phase 1. The sky image was created
from six of the seven images and we iterated only once to produce the stacked image.
%

%
%%%%%%%%%%%%%%%%%%%%%%%%%%%%%%%%%%%%%%%%%%%%
%%%%% Observations: spectroscopy %%%%%
%%%%%%%%%%%%%%%%%%%%%%%%%%%%%%%%%%%%%%%%%%%%
%
%
\subsection{Optical and near-infrared low-resolution spectroscopy}
\label{esdT:Obs_spec}
\subsubsection{GTC OSIRIS}
\label{esdT:Obs_spec_OSIRIS}

We obtained optical spectroscopy of WISE1810 with GTC OSIRIS twice. The observations were conducted in service mode on the night of 6 September 2020 between UT\,=\,21h50m  and 23h25m as part of the DDT programme GTC2020-161 (PI Lodieu) and repeated on 19 September 2020 as part of a second DDT programme GTC2020-163 (PI Lodieu). On the first night the moon was illuminated at 79\,\%, but the elevation was low ($<$20 degrees), the sky clear, and the seeing better than 0\farcs9. We collected three exposures of 1800\,s each with an offset of 4\farcs0 along the slit to optimise the sky subtraction in the red part of the spectrum. We collected another two exposures of 1800\,s each with the same instrumental configuration on the second night between UT\,=\,21h27m and 22h30m (programme GTC2020-163). The sky was clear, dark (without moon), and the seeing was around 0\farcs7.

We observed WISE1810 at parallactic angle with the R1000R grating and a slit of 0\farcs8, providing a resolving power of $R \sim 800$ in the central region. The wavelength coverage of this instrumental set-up is 510--1000 nm, but our target is detected only redwards of $\approx$800 nm. 
A spectro-photometric standard star, Ross\,640 \citep[DZA5.5;][]{Wesemael93} was also observed with the same setting as part of the GTC calibration scheme. The standard star was also observed with the $z$ filter to correct for the second-order contamination of the grating and calibrate our target as far to the red wavelengths (at around 1 $\mu$m) as possible.

\begin{figure}
   \centering
   \includegraphics[width=\linewidth, angle=0]{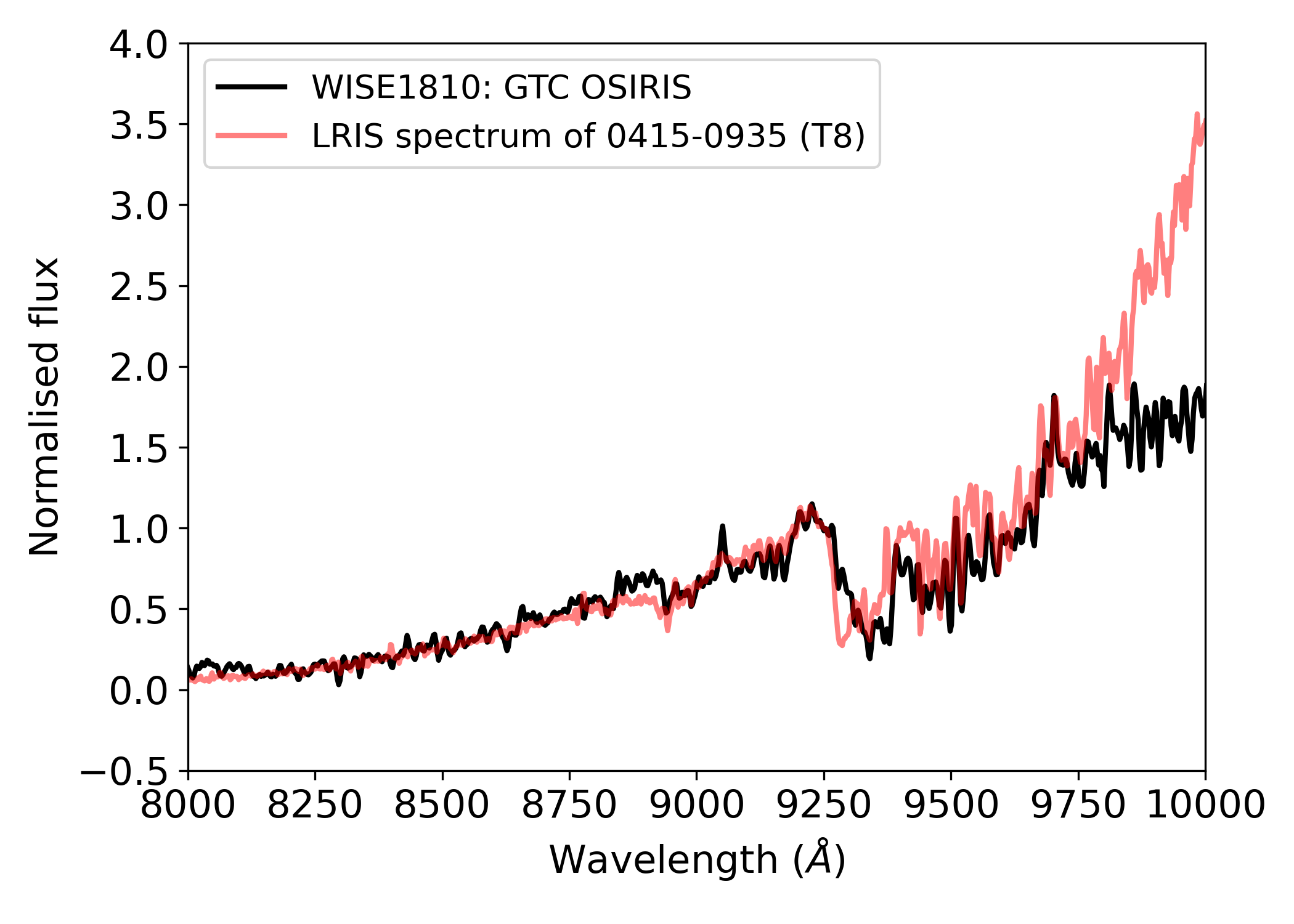}
   \caption{
   GTC OSIRIS low-resolution (R$\sim$800) optical spectrum of WISE1810 normalised at 925 nm.
   Overplotted in red for comparison is a known T0 spectral template, SDSS\,0837$-$0000 \citep[R$\sim$1200][]{burgasser03d,kirkpatrick05}.
   }
   \label{fig_esdT:GTCspectrum}
\end{figure}

We carried out the data reduction of the optical spectra with IRAF\@.
First, we subtracted the raw median-combined bias frame from the raw spectrum and then divided by a normalised dome flat field. Afterwards, we extracted each spectrum separately after subtracting the target frame closer in time and choosing the optimal sky background and an adequate aperture. We calibrated them in wavelength with the Xe-Ne-ThAr arc lamps taken in the afternoon with an $rms$ better than 0.15\,\AA{}. Finally, we averaged all five individual spectra removing the two extreme values. We show the final spectrum in Fig.\ \ref{fig_esdT:GTCspectrum}, compared to the known field T0 spectral template, SDSS\,0837$-$0000 \citep{burgasser03d} overplotted in red. The OSIRIS spectrum of WISE1810 is cut bluewards of 800 nm because there is very little or no flux from the target at these short wavelengths.

%
%%%%%%%%%%%%%%%%%%%%%%%%%%%%%%%%%%%%%%%%%%%%
%%%%% Figure %%%%%
%%%%%%%%%%%%%%%%%%%%%%%%%%%%%%%%%%%%%%%%%%%%
%
%

%
%%%%%%%%%%%%%%%%%%%%%%%%%%%%%%%%%%%%%%%%%%%%
%%%%% Figure: EMIR spectrum %%%%%
%%%%%%%%%%%%%%%%%%%%%%%%%%%%%%%%%%%%%%%%%%%%
%
\begin{figure}
   \centering
   \includegraphics[width=\linewidth, angle=0]{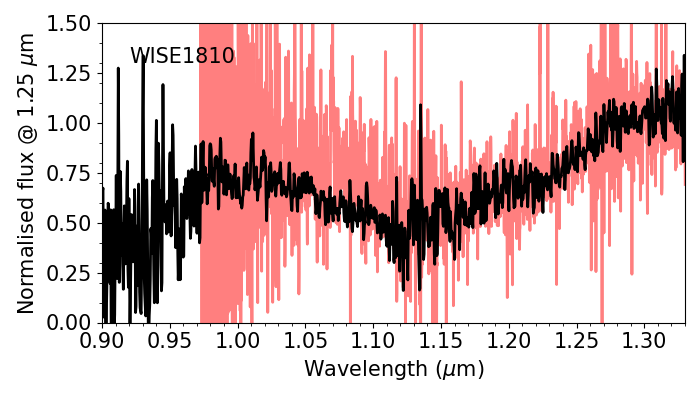}
   \caption{
GTC EMIR $YJ$ spectrum (black) covering 880--1330 nm with a resolution
of about 600\@. The spectrum of \citet{schneider20a} with a resolution of 2600 was plotted beneath the EMIR spectrum to show the difference in quality over the common wavelength range.
   }
      \label{fig_esdT:plot_EMIR_spec}
\end{figure}
\subsubsection{GTC EMIR}
\label{esdT:Obs_spec_EMIR}

We obtained a low-resolution infrared spectrum of WISE1810$-$10 using  the $YJ$ grating of the 
EMIR spectrograph and a slit-width of 1\farcs0. This instrumental set-up yields a wavelength coverage of 880--1330 nm with a resolving power of $R \approx 600$. We created two observing blocks of 2 h on-source time divided into
individual exposures of 360\,s each in an ABBA pattern with offsets of 10\farcs0 along the slit. 
The observations were conducted in service mode on two consecutive nights (2021 June 25 and 26) 
as part of programme GTCMULTIPLE3A-21A (PI Lodieu). A hot, telluric standard star, HIP\, 88374  
\citep[B9V;][]{houk99}, was observed right after the target with on-source integrations of 5\,s 
and one single ABBA pattern with an offset of 52 pixels (or $\sim$10\farcs0) along the slit.

We also observed the standard SDSS\,J134646.45$-$003150.4, a T6 dwarf \citep{burgasser00a} at 14.6 pc
\citep{faherty12} on 7 July 2021 with the same EMIR instrumental set-up to serve as a radial velocity standard.
This dwarf has $v_r$\,=\,$-$17.5$\pm$0.6 km\,s$^{-1}$ \citep{hsu21} and was selected because of its low spectroscopic rotational velocity \citep{zapatero06,hsu21}. The on-source integration time of each individual exposure was 240\,s; the ABBA pattern was repeated to reach a total exposure time of 1920\,s. We also observed an A0V telluric standard star, HD\,116960 ($J$\,=\,7.93 mag), at a similar air mass just after the observations of the T6 dwarf. For the telluric standard, we collected four spectra of 30\,s each in an ABBA pattern.

We combined all individual spectra with an in-house Python code. We created master flat fields combining the five frames provided by the observatory with a median filter. We normalised the flat-field image
and removed it from each individual target frame. We subtracted the two nearest A
and B positions to remove the sky contribution and then averaged the A$-$B and B$-$A images 
without any offset. Afterwards, we measured the offset between the first and second exposure
to combine the A$-$B and B$-$A images into a single image. We ended up with two combined
2D spectra of WISE1810 taken on two distinct nights and one 2D spectrum of the radial velocity standard dwarf. We calibrated the spectra in wavelength with a 
combination of three lamps of xenon, HgAr, and neon.
Finally, we extracted the combined spectra under the IRAF environment.
We repeated the same process with the associated telluric stars.
To correct for the telluric absorption, we divided the averaged spectra of WISE1810 and SDSS\,J134646.45$-$003150.4 by the data of the 
hot stars and multiplied by the black body of B9V and A0V stars downloaded from the ESO website. 

The final spectrum of WISE1810
is overplotted in Figure~\ref{fig_esdT:plot_EMIR_spec} on the spectrum published
by \citet{schneider20a}. The EMIR spectrum has a S/N value about 2--3 times better than the \citet{schneider20a} spectrum; even so, with the exception of strong water vapour bands, no other atomic or molecular features are clearly seen. We can impose an upper limit of 2\,\AA{} on the pseudo-equivalent width of any narrow feature that might be present in the EMIR data, for example th K\,{\sc i} lines at around 1.25 $\mu$m and the VO band at 1.06 $\mu$m. This upper limit is clearly below the pseudo-equivalent widths measured on ultra-cool M, L, and early-T dwarfs (3--10\,\AA{}) but not significantly different from the measurements derived for mid-to-late T dwarfs \citep[$\leq$3\,\AA{}][]{faherty14a}.

%
%%%%%%%%%%%%%%%%%%%%%%%%%%%%%%%%%%%%%%%%%%%%
%%%%% Photometric analysis %%%%%
%%%%%%%%%%%%%%%%%%%%%%%%%%%%%%%%%%%%%%%%%%%%
%
%
\section{Photometric analysis}
\label{phot_analysis}
%

%
%%%%%%%%%%%%%%%%%%%%%%%%%%%%%%%%%%%%%%%%%%%%
%%%%% Table: Photometry
%%%%%%%%%%%%%%%%%%%%%%%%%%%%%%%%%%%%%%%%%%%%
%
\begin{table}
\caption{Photometry of WISE1810. \label{tab:phot} }
\centering
\tiny
\begin{tabular}{@{\hspace{0mm}}l c c c l@{\hspace{0mm}}}
\hline
UT Date  &  JD           & Magnitude & Fil$^a$ & Instrument \\
         & ($-$2450000)  & (mag).    &         &            \\
\hline
2010 Jun 27 & 5374.9601 &   17.264$\pm$0.018$^b$ & $J$  & GPS \\
2013 Aug 31 & 6535.7340 &   18.962$\pm$0.151 & $y$  & PanSTARRS \\
2014 Aug 15 & 6884.7449 &   18.922$\pm$0.107 & $y$  & PanSTARRS \\
2017 Jun 19 & 7923.7030 &   17.240$\pm$0.029 & $J$  & VIRCAM \\
2017 Aug 09 & 7974.6343 &   17.244$\pm$0.044 & $J$  & VIRCAM \\
2020 Aug 18 & 9080.3734 &   20.222$\pm$0.035 & $z$  & ALFOSC \\
2020 Sep 06 & 9099.4053 &   20.117$\pm$0.028 & $z$  & OSIRIS \\
2020 Sep 19 & 9112.3746 &   20.150$\pm$0.048 & $z$  & OSIRIS \\
2020 Sep 28 & 9121.3746 &   20.203$\pm$0.060 & $z$  & OSIRIS \\
2020 Oct 09 & 9132.3490 &   20.087$\pm$0.048 & $z$  & OSIRIS \\
2020 Oct 23 & 9146.3287 &   20.203$\pm$0.059 & $z$  & OSIRIS \\
2020 Oct 25 & 9148.3330 &   20.040$\pm$0.072 & $z$  & OSIRIS \\
2020 Nov 08 & 9162.3189 &   20.209$\pm$0.030 & $z$  & OSIRIS \\
2021 Feb 16 & 9261.7513 &   20.246$\pm$0.038 & $z$  & ALFOSC \\
2021 Mar 04 & 9277.7749 &   17.285$\pm$0.025 & $J$  & EMIR \\
2021 Mar 25 & 9298.6813 &   17.329$\pm$0.048 & $J$  & Omega2000 \\
2021 Apr 21 & 9325.6286 &   17.254$\pm$0.064 & $J$  & Omega2000 \\
2021 Apr 22 & 9326.6889 &   17.276$\pm$0.076 & $J$  & EMIR \\
2021 May 10 & 9344.6292 &   20.035$\pm$0.008 & $z$  & HiPERCAM \\
2021 May 26 & 9360.5666 &   17.220$\pm$0.025 & $J$  & EMIR \\
2021 May 29 & 9363.6984 &   17.303$\pm$0.021 & $J$  & EMIR \\
2021 Jun 22 & 9387.5751 &   17.297$\pm$0.029 & $J$  & EMIR \\
2021 Jun 26 & 9392.4763 &   17.330$\pm$0.080 & $J$  & EMIR \\
2021 Jul 01 & 9396.6155 &   17.360$\pm$0.120 & $J$  & Omega2000 \\
2021 Jul 18 & 9414.4646 &   17.305$\pm$0.025 & $J$  & EMIR \\
2021 Jul 28 & 9423.4985 &   17.672$\pm$0.194 & $J$  & Omega2000 \\
2021 Aug 07 & 9434.3893 &   17.277$\pm$0.038 & $J$  & EMIR \\
2021 Aug 24 & 9451.3927 &   17.345$\pm$0.130 & $J$  & Omega2000 \\ 
2021 Aug 25 & 9452.4396 &   17.291$\pm$0.025 & $J$  & EMIR \\
2021 Oct 19 & 9507.2902 &   17.203$\pm$0.129 & $J$  & Omega2000 \\
\hline
%\noalign{\smallskip}
\multicolumn{5}{c}{Average/adopted photometry} \\
            &           &   23.871$\pm$0.104 & $i$  & HiPERCAM \\
            &           &   20.147$\pm$0.083 & $z$  & HiPERCAM$+$OSIRIS \\
            &           &   18.942$\pm$0.129 & $y$  & PanSTARRS \\
            &           &   17.291$\pm$0.044 & $J$  & EMIR$+$Om2000\\
            &           &   16.516$\pm$0.029 & $H$  & GPS \\
            &           &   17.097$\pm$0.165 & $K$  & GPS \\
            &           &   13.924$\pm$0.033 & $W1$ & NEOWISE \\
            &           &   12.584$\pm$0.038 & $W2$ & NEOWISE \\
            &           &   11.445$\pm$0.219 & $W3$ & WISE \\
%\noalign{\smallskip}
\hline
%\noalign{\smallskip}
%\begin{tablenotes}
\multicolumn{5}{l}{$^a$ $izy$ are in the AB photometric system.}\\
\multicolumn{5}{l}{$^b$ Photometry is taken from the GPS \citep{lucas08} catalog.}
%\end{tablenotes}
\end{tabular}
\end{table}

%%%%%%%%%%%%%%%%%%%%%%%%%%%%%%%%%%%%%%%%%%%%
%%%%% Figure: WISE photometry %%%%%
%%%%%%%%%%%%%%%%%%%%%%%%%%%%%%%%%%%%%%%%%%%%
%
\begin{figure}
   \centering
   \includegraphics[width=\linewidth, angle=0]{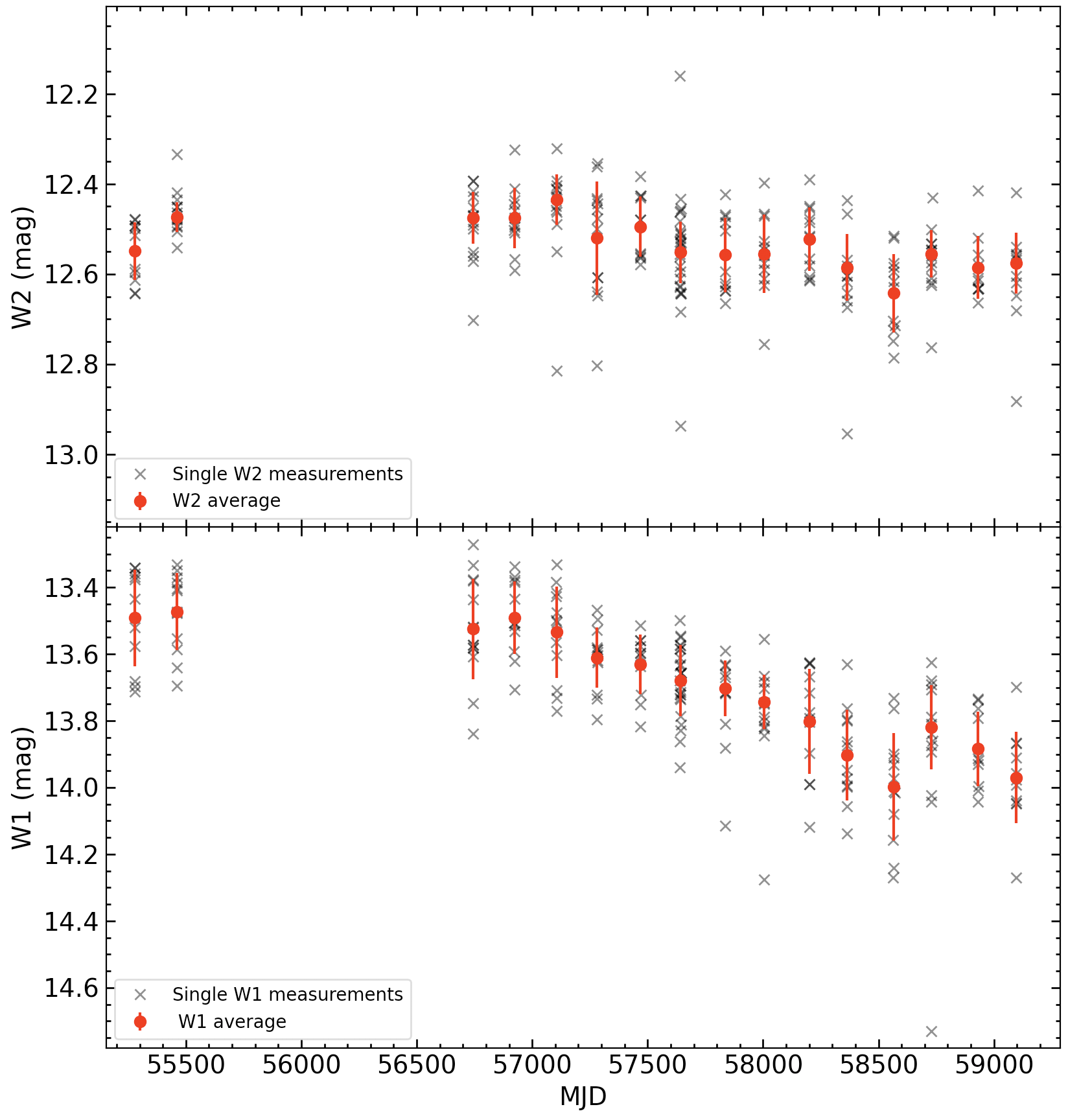}
   \caption{WISE $W2$ {\sl (top)} and $W1$ {\sl (bottom)} photometry of WISE1810 over time. Individual measurements are plotted as gray crosses, the average photometry per observing epoch is plotted as red dots. Error bars account for the photometric dispersion. Both filters show a decreasing trend likely due to flux contamination by a background star. NEOWISE data \citep{mainzer11} were acquired between modified Julian dates (MJD) = 56742.7 and 59096.3.
   }
   \label{fig_esdT:wise}
\end{figure}

The photometric analysis of the multiple datasets was carried out using routines within DAOPHOT in IRAF. Instrumental aperture and point-spread-function (PSF) photometry were obtained by using a circular aperture typically four times larger than the average FWHM of the frames (in the vicinity of our target) and a sky annulus of radius 4\,$\times$\,FWHM and width of 5 pixels. The sky was computed as the mode of the annulus to avoid the flux contribution of nearby sources. The PSF of each individual frame was constructed by selecting from three to five stars that are not blended with other sources in the field near our target and with small error bars in the aperture photometry. The instrumental magnitudes were converted into observed magnitudes using Panoramic Survey Telescope and Rapid Response System \citep[PanSTARRS;][]{chambers16} and 2MASS magnitudes of $J$=14--16.5 mag stars in the field. We made every possible effort to use the same photometric calibrators from epoch to epoch. The typical uncertainty in the calibration of the $z$- and $J$-band images is of the order of 0.02--0.04 mag, which is quadratically added to the PSF errors (given by IRAF) to obtain the final quoted error bars of our photometry provided in Table~\ref{tab:phot}. We applied the same method to extract the $y$-band photometry of WISE1810 using the two late PanSTARRS images (see Section~\ref{esdT:Obs_parallax}). The data are also shown in Table~\ref{tab:phot}.

WISE1810 is clearly detected in the ALFOSC and OSIRIS $z$-band images. It is also detected at the 4--5$\sigma$ level in the $i$-band filter with HiPERCAM. Our derived magnitude is $i = 23.871 \pm 0.104$ (AB system). This is the first time that a metal-poor, ultra-cool dwarf has been detected in the $i$ filter. We do not detect WISE1810 in the bluest bands and derive 3\,$\sigma$ upper limits of $r$\,$\ge$\,25.2 mag and $g$\,$\ge$\,26.6 mag. These limits are more restrictive than those provided for the image around WISE1810 taken by the VLT Survey telescope reported in the ESO science portal archive ($u, g, r, i$\,=\,21.56, 22.79, 22.15, 21.42)\footnote{\url{http://archive.eso.org/scienceportal/home}}. All of our $g, r, i, z$ photometry is calibrated using the PanSTARRS database. 

We downloaded the WISE photometry of WISE1810 from the Infrared Science Archive (IRSA\footnote{https://irsa.ipac.caltech.edu/about.html}); all the available WISE \citep{wright10} and NEOWISE \citep{mainzer11} data amounts to 204 individual measurements per $W1$ and $W2$ filters spanning 10.5 yr of observations. Figure~\ref{fig_esdT:wise} shows the photometric time series where the average photometry per observing epoch is also illustrated. The most striking feature is the long-term decreasing trend of the $W1$ magnitude: WISE1810 gets fainter by about 0.4 mag in $W1$. This behaviour is also present in the $W2$ data, but with a significantly smaller amplitude ($\approx$0.1 mag). We ascribe this feature to the flux contamination by a background star in the field, the same contaminating source mentioned by \citet[their Figure~1]{schneider20a}. Contamination is stronger at the first WISE epochs, thus making WISE1810 brighter. As WISE1810 moves away from the background source due to its high proper motion, contamination becomes less significant. Over the most recent five to six NEOWISE epochs, the $W1$ magnitudes of WISE1810 appear to flatten, implying that flux contamination is less significant. The contaminating source (2MASS\,J18100620$-$1010026) is unresolved in all of our optical and near-infrared data; therefore, it is a likely star of our Galaxy with a $W1-W2$ colour around a null value. This contrasts strongly with the very red $W1-W2$ index of our target, which explains that the $W2$ photometry is less contaminated than the $W1$ data. By averaging the last six NEOWISE epochs, we derived the following photometry for WISE1810: $W1 = 13.924 \pm 0.033$ and $W2 = 12.584 \pm 0.038$ mag ($W1 - W2 = 1.340 \pm 0.050$ mag), where the error bars correspond to the standard error of the mean. These values are included at the bottom of Table~\ref{tab:phot} for completeness. They are clearly fainter by $\approx$0.3 ($W1$) and $\approx$0.1 ($W2$) mag than those from \citet{schneider20a}.

%
%%%%%%%%%%%%%%%%%%%%%%%%%%%%%%%%%%%%%%%%%%%%
%%%%% parallax %%%%%
%%%%%%%%%%%%%%%%%%%%%%%%%%%%%%%%%%%%%%%%%%%%
%
%
\section{Astrometric analysis}
\label{esdT:Obs_parallax}
\subsection{Astrometry}

To derive the trigonometric parallax and proper motion of WISE1810, we used all the recent imaging data presented in Section~\ref{esdT:Obs_phot} plus the following older data: 
\begin{itemize}
    \item The Galactic Plane Survey $J$-band image (GPS; \citealt{lucas08}), which is part of the UKIRT Infrared Deep Sky Survey (UKIDSS; \citealt{lawrence07}). The GPS data were acquired on 2010 Jun 27. WISE1810 was also detected twice in the same year by the {\sl WISE} survey \citep{wright10}; however, as stated in \citet{schneider20a}, WISE1810 is blended with a background object in these earliest {\sl WISE} epochs. For this reason, we did not use the astrometry from the {\sl WISE} All-Sky data in our analysis. 
    \item The $y$-band images from the PanSTARRS first data release \citep{chambers16} taken on 14 August 2011, 31 August 2013, and 15 August 2014\@. We used the $y$-band data because WISE1810 is brighter at these wavelengths than at the other blue PanSTARRS filters. WISE1810 appears slightly blended with another background source in the first PanSTARRS epoch.
    \item The eXtended VISTA Variables in the Via Lactea $J$-band images (VVVX; \citealt{minniti10,saito12}) taken with the VISTA InfraRed CAMera (VIRCAM) instrument on 19 June 2017 and 9 August 2017. We employed the $J$-band images for consistency with the GPS, EMIR, and Omega2000 data. We also obtained the $J$-band photometry of the VVVX images (Table~\ref{tab:phot}) following the same methodology as that described in the previous section.
\end{itemize}
None of the PanSTARRS and VIRCAM data were included in the discovery paper \citep{schneider20a}. However, all of these images are useful for the proper motion determination and, in the case of the VIRCAM data, their quality is sufficiently good for the parallax measurement.

We employed the oldest GPS observations as the fundamental reference frames to which all other PanSTARRS, VIRCAM, ALFOSC, OSIRIS, EMIR, Omega2000, and HiPERCAM images are compared for the following reasons: they are the first data available for WISE1810, the images have an adequate pixel size and excellent seeing, the target is not blended with any other source and is detected with significant S/N (see the left panel of Fig.~\ref{fig_esdT:@ISE1810images}).
Using the {\sl daofind} command within IRAF we identified all sources with photon peaks with detection above 8--10 $\sigma$ in all images, where $\sigma$ stands for the noise of the background, and full width at half maximum (FWHM) resembling that of unresolved objects (i.e., extended sources were mostly avoided). In addition, we ensured that the detected sources lay within the linear regime of the detector's response. The centroids of detected objects were computed by estimating the $x$ and $y$ pixel positions of the best fitting one-dimensional Gaussian functions on each axis; typical associated errors are about 3--5\,\%~of a pixel or better. WISE1810 appears rather faint or slightly blended in some images (see below), therefore, its errors in $x$ and $y$ coordinates can exceed the 5\,\%~uncertainty. For the astrometric analysis, we used a field of view of 1.1\,$\times$\,1.1 arcmin$^2$ centred around WISE1810. In this area, the number of sources identified per frame typically exceeded 100 objects. 

Pixel ($x$, $y$) coordinates were transformed between different epochs using the {\sl geomap} routine within IRAF, which applied a polynomial of third order in $x$ and $y$, and computed linear terms and distortions terms separately. The linear term included an $x$ and $y$ shift and an $x$ and $y$ scale factor, a rotation, and a skew. The distortion surface term consisted of a polynomial fit to the residuals of the linear term. The ($x$, $y$) astrometric transformation between observing epochs and the reference epoch was an iterative step, that included the rejection of objects deviating by more than 1.8--2\,$\sigma$, where $\sigma$ corresponds to the dispersion of the transformation. The typical dispersion of the coordinates transformation was always 10\,\%~of a pixel or better. We scaled all pixels to the size of the OSIRIS pixel, which was well determined at $129.11 \pm 0.18$ mas by \citet{sahlmann16}. Since we observed with the standard mode of OSIRIS, we used the $2\times2$ binning, thus the OSIRIS pixel size in our analysis is twice the value published in \citet{sahlmann16}. The final values of the different instruments pixel sizes are provided in Table~\ref{tab_esdT:pixels}. With the only exception of ALFOSC, we do not observe any significant difference between the $x$ and $y$ directions, that is, the pixels appear to be squared. All values are compatible at the level of 2\,$\sigma$ with the nominal pixel sizes given in the instrumental manuals or published elsewhere (here, $\sigma$ stands for the quoted uncertainty associated with each instrument pixel size). There are only three exceptions: PanSTARRS has a plate scale of 258 mas according to \citet{chambers16}, EMIR has a pixel size of 200 mas according to the instrument manual, and ALFOSC $y$-axis projects onto 213.8 mas on the sky according to the instrument manual; however, our determination for the $x$-axis is compatible with the nominal value. In all three cases, our derivations are about 3\,\%~smaller. 

Because our data were acquired at red optical and near-infrared wavelengths
(mostly 0.84--1.33 $\mu$m) and given the relatively small field of view used for the coordinates transformation, corrections by refraction due to the Earth’s atmosphere are expected to be small \citep{filippenko82}. Furthermore, since we are using relative astrometry, only the differential refraction is relevant, and this effect is in practice accounted for by using polynomial astrometric transformations of degree three and higher (see also \citealt{fritz10}). Regarding the chromatic differential refraction, our target is redder and shows stronger water vapour absorption at near-infrared wavelengths than the vast majority of the reference sources used in the astrometric transformations. Nevertheless, various works in the literature \citep{monet92, faherty12} have demonstrated that the differential colour refraction corrections are minimal at infrared wavelengths. Using \citet{stone02}, who computed ranges in differential colour refraction for a zenith distance of 60 deg as a function of wavelength, we estimated them to be about a few milliarcseconds for the largest zenith distances of our observations, which are smaller than the quoted astrometric uncertainties for individual images. For small zenith distances, the corrections are of the order of sub-milliarcseconds. Therefore, we did not attempt to apply the differential chromatic refraction correction\footnote{We caution, however, that differential chromatic refraction can be a source of systematic errors in astrometric studies. to the ($x$, $y$) positions in our astrometric analysis procedure.}

In Table~\ref{tab_esdT:tab_astrometry} we provide the differential astrometry of WISE1810 as a function of the observing date: the (d$x$, d$y$) pixel data were converted into projected separations in right ascension (d$\alpha$\,cos\,$\delta$) and declination (d\,$\delta$) using the plate scales reported in Table~\ref{tab_esdT:pixels}. Error bars were computed as the quadratic sum of the dispersion of the pixel coordinate transformations and the uncertainties of the centroid determination of WISE1810 for each frame. The PanSTARRS and Omega2000 data show the largest astrometric error bars due to the faint detection of our target in the PanSTARRS images and the large pixel size of Omega2000. WISE1810 appears slightly blended along the $y$-axis (declination) with other background sources in the field in the most recent Omega2000 images.

%
%%%%%%%%%%%%%%%%%%%%%%%%%%%%%%%%%%%%%
%%%%% Table: plate scales %%%%%
%%%%%%%%%%%%%%%%%%%%%%%%%%%%%%%%%%%%%
%
\begin{table}
\caption{Detector pixel size used in the astrometric analysis. \label{tab_esdT:pixels} }
\centering
\tiny
\begin{tabular}{@{\hspace{-0.1mm}}l @{\hspace{2mm}}c @{\hspace{0.6mm}}l@{\hspace{0.1mm}} }
\hline
\noalign{\smallskip}
Instrument  &  Pixel size & Reference \\
            & (mas)       &  \\
\noalign{\smallskip}
\hline
\noalign{\smallskip}
OSIRIS (unbinned) & 129.11$\pm$0.18 & \citet{sahlmann16} \\
GPS               & 201.22$\pm$0.38 & This paper \\
ALFOSC            & 214.02 ($x$), 207.40 ($y$)$\pm$0.52 & This paper \\
PanSTARRS         & 250.1$\pm$0.2 & This paper \\
VIRCAM            & 341.7$\pm$0.3 & This paper \\
EMIR              & 194.5$\pm$0.2     & This paper \\
Omega2000         & 449.45$\pm$0.45     & This paper \\
HiPERCAM          & 80.67$\pm$0.08     & This paper \\
\noalign{\smallskip}
\hline
\end{tabular}
\end{table}
%

%
%%%%%%%%%%%%%%%%%%%%%%%%%%%%%%%%%%%%%
%%%%% Table: Astrometry for WISE1810 %%%%%
%%%%%%%%%%%%%%%%%%%%%%%%%%%%%%%%%%%%%
%
\begin{table}
\caption{Relative astrometry of WISE1810. \label{tab_esdT:tab_astrometry} }
\centering
\tiny
\begin{tabular}{@{\hspace{-0.1mm}}l @{\hspace{2mm}}l @{\hspace{0.6mm}}r @{\hspace{1mm}}r @{\hspace{1mm}}c @{\hspace{1mm}}l}
\hline
\noalign{\smallskip}
UT Date  &  JD           & \multicolumn{1}{c}{d\,$\alpha$\,cos\,$\delta$}  & \multicolumn{1}{c}{d\,$\delta$} & Filt & Instrument \\
         & ($-$2450000)  & \multicolumn{1}{c}{(arcsec)}     & \multicolumn{1}{c}{(arcsec)}    &      &            \\
\noalign{\smallskip}
\hline
\noalign{\smallskip}
2010 Jun 27 & 5374.9601 &     0.00$\pm$0.05 &    0.00$\pm$0.05 & $J$  & GPS \\
2011 Aug 14 & 5787.8377 &  $-$1.21$\pm$0.13 & $-$0.59$\pm$0.21 & $y$  & PanSTARRS \\
2013 Aug 31 & 6535.7340 &  $-$3.46$\pm$0.11 & $-$0.87$\pm$0.10 & $y$  & PanSTARRS \\
2014 Aug 15 & 6884.7449 &  $-$4.32$\pm$0.06 & $-$0.96$\pm$0.10 & $y$  & PanSTARRS \\
2017 Jun 19 & 7923.7030 &  $-$7.16$\pm$0.04 & $-$1.68$\pm$0.03 & $J$  & VIRCAM \\
2017 Aug 09 & 7974.6343 &  $-$7.40$\pm$0.03 & $-$1.68$\pm$0.03 & $J$  & VIRCAM \\
2020 Aug 18 & 9080.3734 & $-$10.46$\pm$0.02 & $-$2.43$\pm$0.05 & $z$  & ALFOSC \\
2020 Sep 06 & 9099.4053 & $-$10.62$\pm$0.02 & $-$2.44$\pm$0.02 & $z$  & OSIRIS \\
2020 Sep 19 & 9112.3746 & $-$10.62$\pm$0.01 & $-$2.55$\pm$0.02 & $z$  & OSIRIS \\
2020 Sep 28 & 9121.3746 & $-$10.66$\pm$0.03 & $-$2.50$\pm$0.02 & $z$  & OSIRIS \\
2020 Oct 09 & 9132.3490 & $-$10.68$\pm$0.02 & $-$2.52$\pm$0.02 & $z$  & OSIRIS \\
2020 Oct 25 & 9148.3330 & $-$10.68$\pm$0.04 & $-$2.57$\pm$0.04 & $z$  & OSIRIS \\
2020 Nov 08 & 9162.3189 & $-$10.77$\pm$0.02 & $-$2.56$\pm$0.02 & $z$  & OSIRIS \\
2021 Feb 16 & 9261.7513 & $-$10.82$\pm$0.04 & $-$2.56$\pm$0.03 & $z$  & ALFOSC \\
2021 Mar 04 & 9277.7749 & $-$10.86$\pm$0.02 & $-$2.60$\pm$0.03 & $J$  & EMIR \\
2021 Mar 25 & 9298.6813 & $-$10.95$\pm$0.05 & $-$2.64$\pm$0.06 & $J$  & Omega2000 \\
2021 Apr 21 & 9325.6286 & $-$11.01$\pm$0.05 & $-$2.70$\pm$0.05 & $J$  & Omega2000 \\
2021 Apr 22 & 9326.6889 & $-$11.01$\pm$0.02 & $-$2.71$\pm$0.02 & $J$  & EMIR \\
2021 May 10 & 9344.6292 & $-$11.09$\pm$0.02 & $-$2.66$\pm$0.02 & $z$  & HiPERCAM \\
2021 May 26 & 9360.5666 & $-$11.16$\pm$0.04 & $-$2.69$\pm$0.06 & $J$  & EMIR \\
2021 May 29 & 9363.6984 & $-$11.18$\pm$0.02 & $-$2.66$\pm$0.02 & $J$  & EMIR \\
2021 Jun 22 & 9387.5751 & $-$11.28$\pm$0.03 & $-$2.68$\pm$0.04 & $J$  & EMIR \\
2021 Jun 26 & 9392.4763 & $-$11.30$\pm$0.02 & $-$2.67$\pm$0.02 & $J$  & EMIR \\
2021 Jul 01 & 9396.6155 & $-$11.40$\pm$0.11 & $-$2.62$\pm$0.12 & $J$  & Omega2000 \\
2021 Jul 18 & 9414.4646 & $-$11.40$\pm$0.03 & $-$2.71$\pm$0.03 & $J$  & EMIR \\
2021 Jul 28 & 9423.4985 & $-$11.44$\pm$0.05 & $-$2.66$\pm$0.14 & $J$  & Omega2000 \\
2021 Aug 07 & 9434.3893 & $-$11.50$\pm$0.03 & $-$2.75$\pm$0.03 & $J$  & EMIR \\
2021 Aug 24 & 9451.3927 & $-$11.55$\pm$0.05 & $-$2.69$\pm$0.06 & $J$  & Omega2000 \\ 
2021 Aug 25 & 9452.4396 & $-$11.56$\pm$0.03 & $-$2.74$\pm$0.03 & $J$  & EMIR \\
2021 Oct 19 & 9507.2902 & $-$11.71$\pm$0.05 & $-$2.90$\pm$0.07 & $J$  & Omega2000 \\
\noalign{\smallskip}
\hline
\end{tabular}
\end{table}
%

%%%%%%%%%%%%%%%%%%%%%%%%%%%%%%%%%%%%%%%%%%%%
%%%%% Figure: parallax %%%%%
%%%%%%%%%%%%%%%%%%%%%%%%%%%%%%%%%%%%%%%%%%%%
%
\begin{figure}
   \centering
   \includegraphics[width=\linewidth, angle=0]{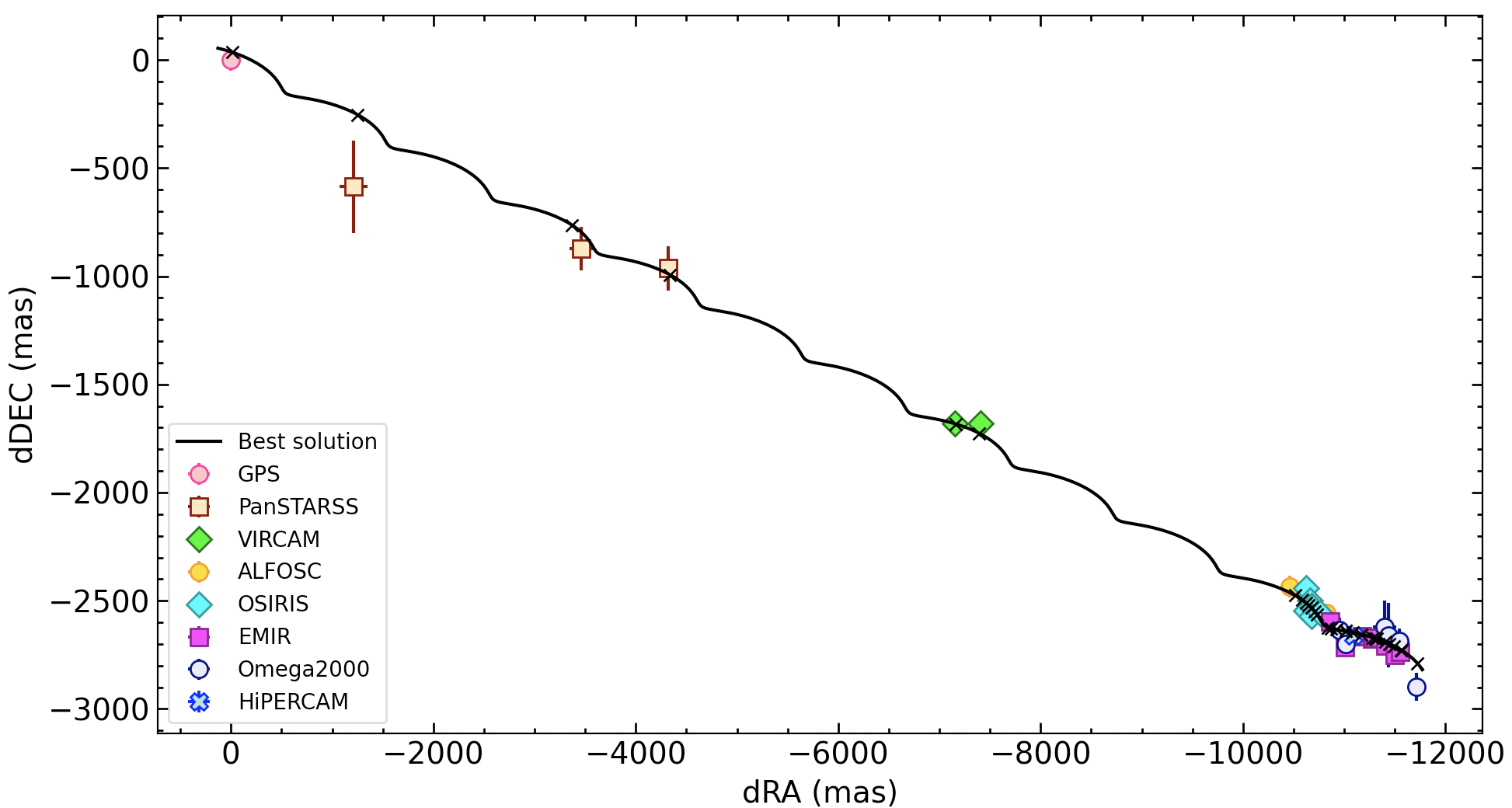}
   \includegraphics[width=\linewidth, angle=0]{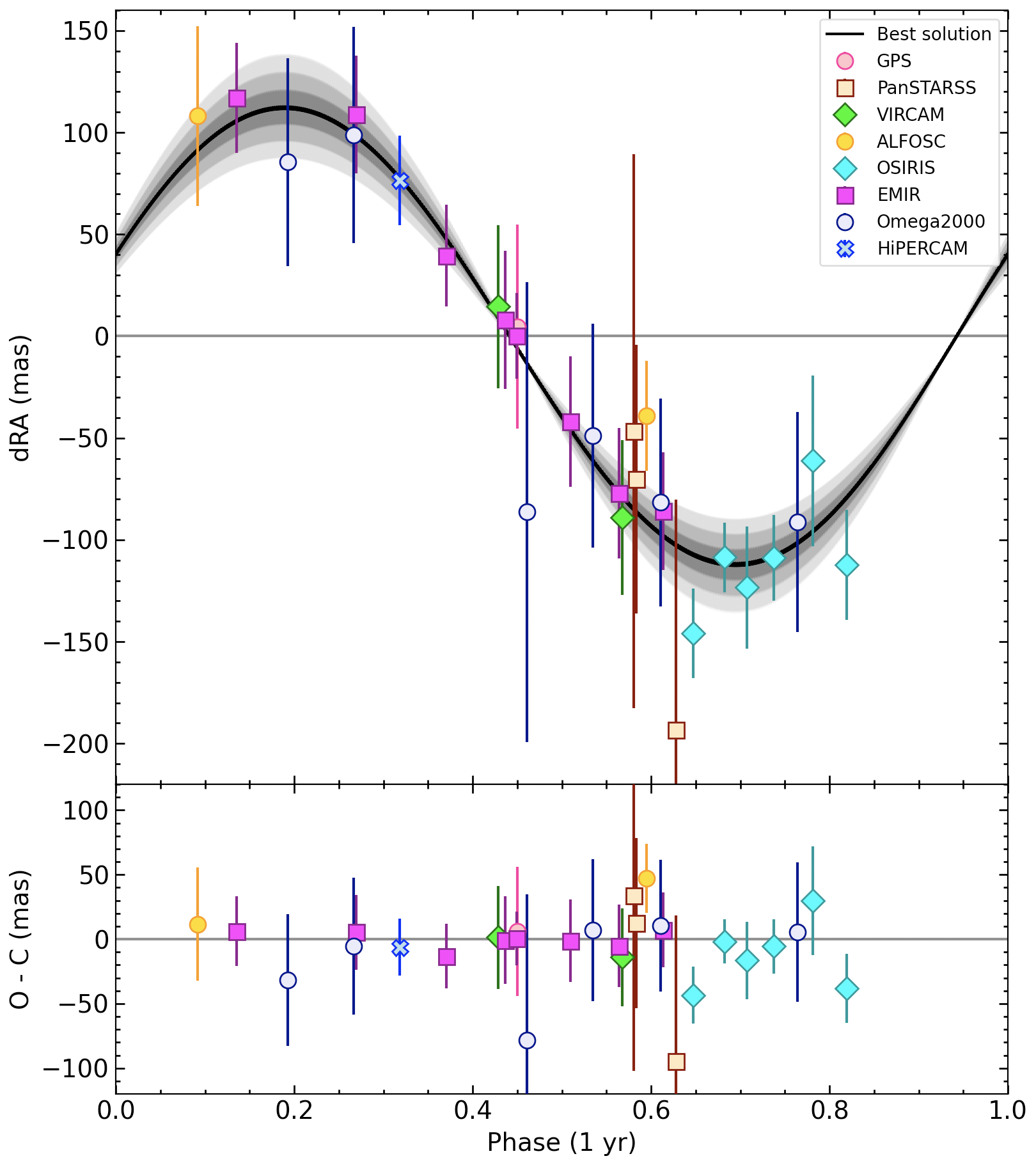}
   \caption{Astrometry of WISE1810\@. {\sl (Top)} Observed relative astrometry of WISE1810 (see legend for the different instruments employed in our study). The GPS data (the oldest images) act as the reference epoch. The best solution including proper motion and trigonometric parallax is shown with a black, solid line. The crosses plotted on the best solution indicate the predicted position of WISE1810 for each observing epoch. {\sl (Bottom)} Relative astrometry (right ascension) after removing the proper motion from the data is folded in phase with the 1 yr period (top subpanel). The amplitude of the variation is the trigonometric parallax. The best parallax solution is shown with the black, solid line and the 1-, 2-, and 3$\sigma$ uncertainty is plotted as a gray-shaded area of decreasing intensity (darkest for the 1$\sigma$). The bottom subpanel illustrates the observed minus computed right ascension residuals.
   }
   \label{fig_esdT:ra_dec}
\end{figure}
\subsection{Parallax and proper motion}

The apparent trajectory of WISE1810 on the sky over the period of 11.3 yr of observations is shown in the top panel of Figure~\ref{fig_esdT:ra_dec}. The (0, 0) astrometric point corresponds to GPS reference data, which happens to be the earliest epoch. As already pointed out by \citet{schneider20a}, WISE1810 moves towards the south-west. The most recent observations in 2020 and 2021 reveal a wobble in the data sequence, which is ascribed to the parallax effect. To derive the proper motion and the trigonometric parallax, we fit the following equations:
\begin{equation}
\label{eq1}
    {\rm d}\alpha \, {\rm cos}\,\delta = \mu_\alpha \, {\rm cos}\,\delta \, (t-t_o) + \pi \, (f_{t}^{\alpha} - f_{o}^{\alpha}) + k_\alpha
\end{equation}
\begin{equation}
\label{eq2}
    {\rm d}\delta = \mu_\delta \, (t-t_o) + \pi \, (f_{t}^{\delta} - f_{o}^{\delta}) + k_\delta
\end{equation}
Here the subscript $o$ indicates the reference epoch, and $f^\alpha$ and $f^\delta$ stand for the parallax factors in right ascension and declination, respectively. The observed parallax is represented by the $\pi$ term, and $\mu_\alpha \, {\rm cos}\,\delta$ and $\mu_\delta$ stand for the proper motion in right ascension and declination. In our analysis we allowed for small offsets in both axes ($k_\alpha$ and $k_\delta$) that would help find statistically better solutions. All the astrometric quantities are given in milliarcseconds and the times $t$ and $t_0$ are measured in Julian days. The parallax factors were computed by following the equations given in \citet{green85} and obtaining the Earth barycentre from the DE405 Ephemeris\footnote{http://ssd.jpl.nasa.gov}.

%
%%%%%%%%%%%%%%%%%%%%%%%%%%%%%%%%%%%%%
%%%%% Table: astrometric priors and solution %%%%%
%%%%%%%%%%%%%%%%%%%%%%%%%%%%%%%%%%%%%
%
\begin{table}
\caption{Priors and posteriors of the astrometric parameters. 
}
\centering
\begin{tabular}{llll}
\hline
\noalign{\smallskip}
Parameter  &  Unit & Prior  & Posterior \\
\noalign{\smallskip}
\hline
\noalign{\smallskip}
$\pi$                       & mas            & $\mathcal{U}$ (0.01, 2000) & 112.2 $^{+8.1}_{-8.0}$ \\
$\mu_\alpha$\,cos\,$\delta$ & mas\,yr$^{-1}$ & $\mathcal{N}$ ($-$1100,  200) & $-$1027.0 $\pm$ 3.5 \\ 
$\mu_\delta$                & mas\,yr$^{-1}$ & $\mathcal{N}$ ($-$200, 100) & $-$246.4 $\pm$ 3.6 \\
$k_\alpha$                  & mas            & $\mathcal{U}$ ($-$9000, 9000) & $-$9.9 $\pm$ 37.1 \\
$k_\delta$                  & mas            & $\mathcal{U}$ ($-$9000, 9000) & 36.2 $\pm$ 37.2 \\
jitter $\alpha$             & mas            & $\mathcal{N}$ (33, 100) & 0$^*$ \\
jitter $\delta$             & mas            & $\mathcal{N}$ (35, 100) & 0$^*$ \\
\noalign{\smallskip}
\hline
\noalign{\smallskip}
\multicolumn{4}{l}{{\tiny $^*$ Smaller than the median of the error bars, thus compatible with a null value.}}
\label{tab_esdT:priors} 
\end{tabular}
\end{table}

%%%%%%%%%%%%%%%%%%%%%%%%%%%%%%%%%%%%%%%%%%%%
%%%%% Figure: corner plot parallax %%%%%
%%%%%%%%%%%%%%%%%%%%%%%%%%%%%%%%%%%%%%%%%%%%
%
\begin{figure}
   \centering
   \includegraphics[width=\linewidth, angle=0]{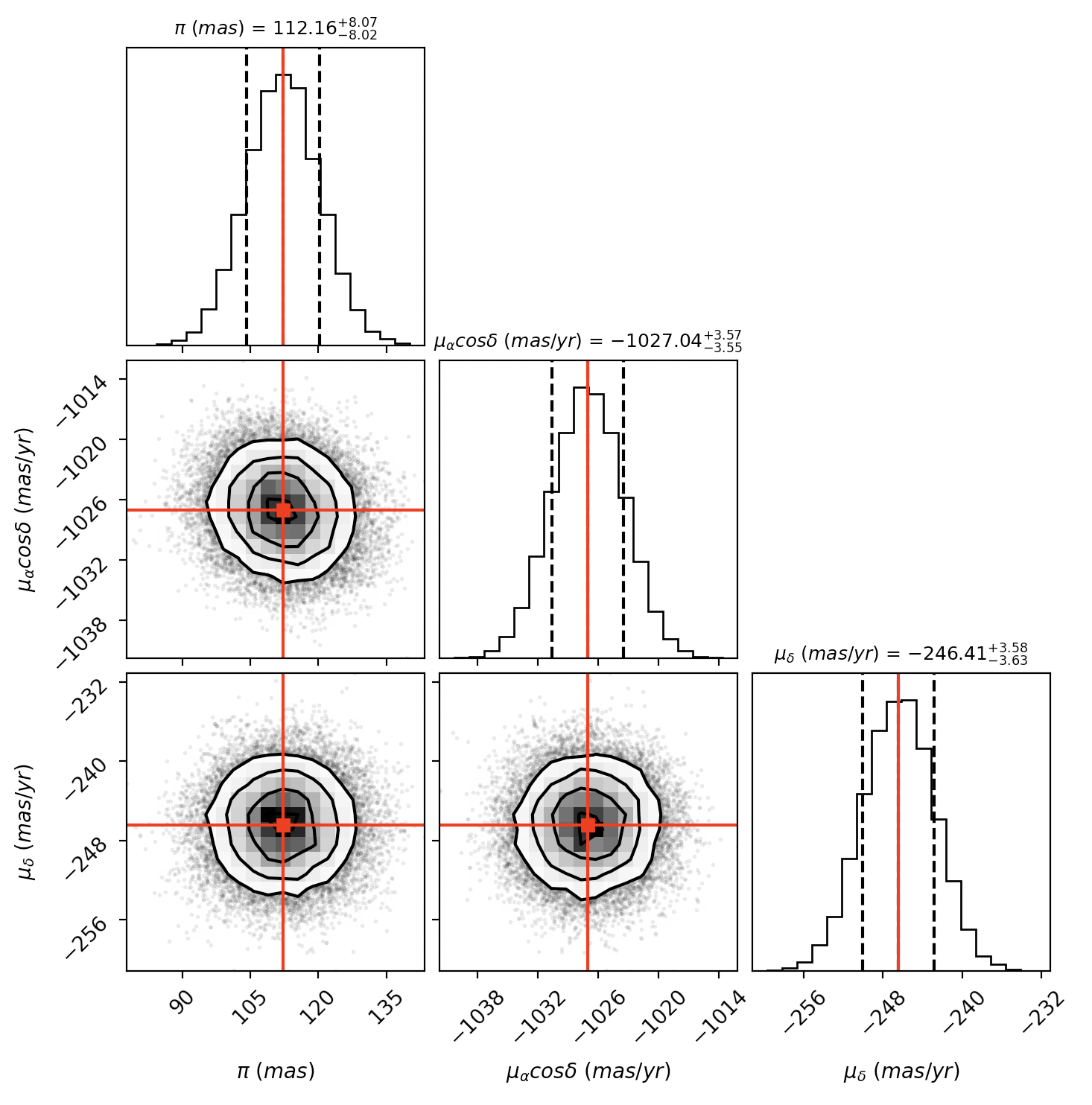}
   \caption{Corner plot displaying the posterior distributions of the  trigonometric parallax and proper motion of WISE1810. The most likely values are indicated.
   }
   \label{fig_esdT:corner}
\end{figure}

We solved Eqs.~\ref{eq1} and~\ref{eq2} simultaneously using Markov chain Monte Carlo methods for sampling the probabilistic distributions of the WISE1810 proper motion, parallax, and offsets. A global jitter term on right ascension and declination was also added to account for possibly underestimated astrometric error bars. We used the Python {\tt PyMC3}  package \citep{salvatier16} to define the model and run the MCMC simulations. Priors on the proper motion were set to normal and centred at the original measurements by \citet{schneider20a} whereas priors on the parallax were set to linear with a wide range of test values. Priors on the jitter terms were set to normal at the median values of the right ascension and declination error bars with a width about three times as big. All priors are summarised in Table~\ref{tab_esdT:priors}. We ran {\tt PyMC3} with a number of chains (14) equal to twice the number of free parameters. After removing 2000 burn-in' steps, we performed 2000 samples per chain. A convergence was reached; it was examined by visually inspecting the stability of the chains and the samples. We confirmed that all chains peak and that the distribution of the samples per chain is flat around the mean posterior values. The medians of the posterior distributions and their $\pm$34.13\,\%~intervals were evaluated and were taken as the final parameters and associated 1-$\sigma$ uncertainties, respectively. The adopted final values and their errors are listed in the last column of Table~\ref{tab_esdT:priors}. The corner plot of the posteriors distributions of the parallax and proper motion is shown in Figure~\ref{fig_esdT:corner}. The adopted solution is plotted together with the observations in Figure~\ref{fig_esdT:ra_dec}. We do not show the declination axis in the bottom panel of Figure~\ref{fig_esdT:ra_dec} for two main reasons. First, the amplitude of the parallactic factor is 4.3 times larger for the right ascension axis, which implies that the parallax is better constrained by the right ascension measurements, and second, in the most recent epochs, WISE1810 lies close to two background sources located to the north and south of our target, thus making our astrometry more uncertain in the declination axis. The mean value of the right ascension residuals (shown in the bottom panel of Figure~\ref{fig_esdT:ra_dec}) varies from instrument to instrument: $-0.2$ mas (EMIR), $-$6.2 mas (HiPERCAM), $-$6.4 mas (VIRCAM), 6.0 mas (GPS), $-$12.6 mas (OSIRIS), $-$16.3 mas (PanSTARRS), $-$15.5 mas (Omega2000), and 29.4 mas (ALFOSC). The dispersion of these residuals for the instruments with three or more measurements are as follows: 6.3 mas (EMIR), 24.4 mas (OSIRIS), 31.3 mas (Omega2000), and 56.2 mas (PanSTARRS). 

Another method for measuring proper motions independently of parallax is to compare the astrometry of WISE1810 taken an exact number of years apart, that is, measurements obtained at approximately the same time of the year (within a few days). Any displacements due to parallax will be minimal; they will be small compared with the large motion of the target and can be ignored. We checked that the WISE1810 proper motion obtained from the two methods agree with each other within the quoted errors. 

Our reference objects are mostly stars in the Galaxy, each with its own distance and proper motion. This introduces a systematic error in the parallax and proper motion determination that must be considered. On the one hand, we expect that the motions of the reference objects are randomly oriented; therefore, their effect will be reduced, and we did not correct for it. On the other hand, the finite distances  to our reference objects diminish part of the true parallax of WISE1810. To convert our relative parallax, $\pi$, into its absolute value, $\omega$, we used the Besançon models \citep{robin03} to simulate\footnote{https://model.obs-besancon.fr/} the population of stars with $J$-band magnitudes between 7 and 17 mag in the 1.1 $\times$ 1.1 and 3 $\times$ 3 arcmin$^2$ fields centred at WISE1810. Simulated stars were allowed to have all possible spectral types and progressive distances between 0 and 10 kpc. We adopted the median of the distribution of the simulated objects distances, 0.2--0.3 mas, as the correction to be added to the relative parallax that comes directly from our fit to obtain the absolute parallax. We note, however, that the correction is significantly smaller than the quoted error bar of the measured relative parallax; therefore, it has very little impact on our results. The final parallax is $\omega = 112.5 ^{+8.1}_{-8.0}$ mas, which translates into a trigonometric distance of $d = 8.9\,^{+0.7}_{-0.6}$ pc.

Our proper motion is significantly smaller in right ascension and larger in declination than the determination of \citet{schneider20a}; these authors mostly use CatWISE data \citep{eisenhardt20} for the astrometric analysis. As acknowledged by \citet{schneider20a}, WISE1810 is likely blended with various sources in mostly all epochs, which clearly contaminates the kinematics study, while our study uses ground-based seeing-limited data where WISE1810 is blended in fewer epochs. Regarding the distance, and based on photometric estimates using absolute magnitudes of early-T dwarfs, \citet{schneider20a} obtained $\sim$14 pc (using the $W2$ magnitude) and $\sim$67 pc (using the $K$ magnitude). According to our results, WISE1810 is located at a much shorter distance than the $K$-band photometric distance and it is about 57\,\%~closer than indicated by the $W2$ photometric distance, which clearly indicates that this object is very peculiar and its photometry looks nothing like an early-T dwarf.

%
%%%%%%%%%%%%%%%%%%%%%%%%%%%%%%%%%%%%%%%%%%%%
%%%%% Properties of WISE1810 %%%%%
%%%%%%%%%%%%%%%%%%%%%%%%%%%%%%%%%%%%%%%%%%%%
%
\section{Properties of WISE1810}
\label{esdT:properties}
\subsection{Photometric variability}
\label{esdT:properties_phot_variability}

\begin{figure}
   \centering
   \includegraphics[width=\linewidth, angle=0]{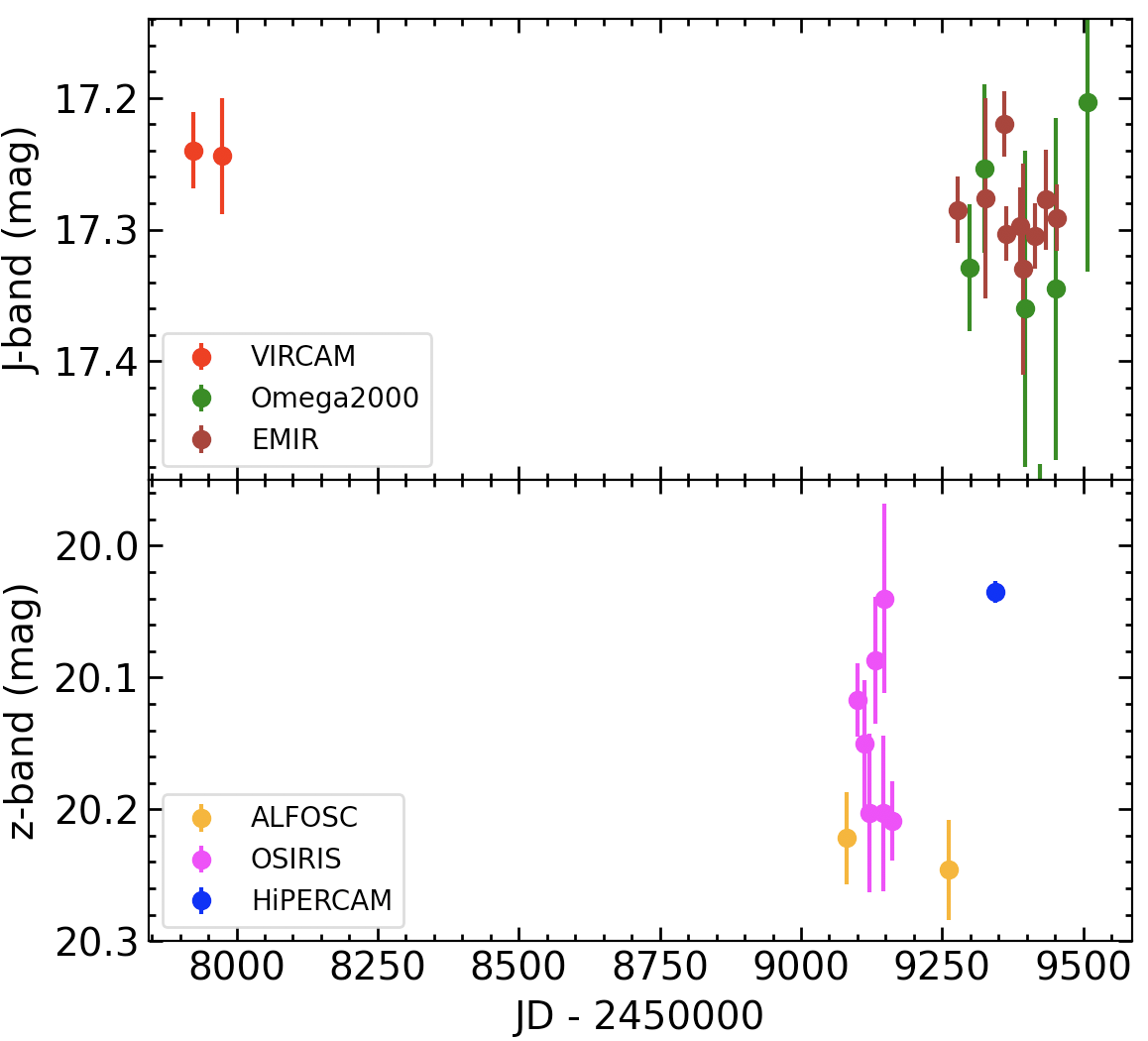}
   \caption{
   Multi-epoch photometry of WISE1810 in the $J$ (top) and $z$ (bottom) filters using our own photometry. There is one deviant $J$-band data point that lies outside the limits of the top panel. Both panels have the same scale on the magnitude axis.
   }
   \label{fig_esdT:phot_var}
\end{figure}

The multiple imaging epochs obtained in the $z$- and $J$-band filters allowed us to investigate the long-term photometric variability of WISE1810 on timescales of months. Figure~\ref{fig_esdT:phot_var} illustrates the photometric sequences in the two filters. The average $J$ magnitude (Vega system) using the Omega2000 and EMIR data is $J$\,=\,17.291 $\pm$ 0.044 mag (where the error bar stands for the dispersion of 14 independent measurements; the standard error of the mean is 0.012 mag). The most discrepant Omega2000 measurement, also affected by the largest uncertainty, is not included in the computation of the mean $J$ magnitude and is not shown in Figure~\ref{fig_esdT:phot_var}. The $J$-band data are based on the 2MASS photometric system. Similarly, at bluer wavelengths, the OSIRIS and HiPERCAM average $z$\,=\,20.147 $\pm$ 0.083 mag is based on the PanSTARRS AB-magnitude system after averaging eight independent measurements. The dispersion of the $z$-band data is larger than that of the $J$-filter because of the intrinsic faintness of WISE1810 at blue wavelengths. 

The recent mean $J$-band magnitude does not deviate from previous measurements provided by the GPS \citep{lucas08} catalogue (Table~\ref{tab:phot}) taken 10--11 yr earlier than our Omega2000 and EMIR data. In addition, the mean $J$-band determination is compatible with the photometry we performed on the VVVX catalog images \citep{minniti10,saito12} collected about 4--5 yr before EMIR. All measurements are compatible at the 1\,$\sigma$ level (where $\sigma$ stands for the photometric errors). Additionally, we do not observe any patterns (Figure~\ref{fig_esdT:phot_var}) and no obvious periodicity is found after the analysis of the Lomb-Scargle periodograms of the $z$- and $J$-band light curves. Therefore, we conclude that WISE1810 does not show any photometric variability with amplitudes larger than 0.13 mag ($J$) and 0.25 mag ($z$) with a confidence of 3\,$\sigma$. 

Taking advantage of the good seeing of some of the optical and near-infrared images, we determined that WISE1810 is not resolved and that any fainter companion with a magnitude difference of $\sim$1.8 mag (in $z$ and $J$) at separations larger than $\approx$5.5 au can be discarded.

%
%%%%%%%%%%%%%%%%%%%%%%%%%%%%%%%%%%%%%%%%%%%%
%%%%% Figure: Photometry vs time %%%%%
%%%%%%%%%%%%%%%%%%%%%%%%%%%%%%%%%%%%%%%%%%%%
% 
%

%
%%%%%%%%%%%%%%%%%%%%%%%%%%%%%%%%%%%%%%%%%%%%
%%%%% Figure: spectral energy distribution SED
%%%%%%%%%%%%%%%%%%%%%%%%%%%%%%%%%%%%%%%%%%%%
% 
%
\begin{figure}
   \centering
   \includegraphics[width=\linewidth, angle=0]{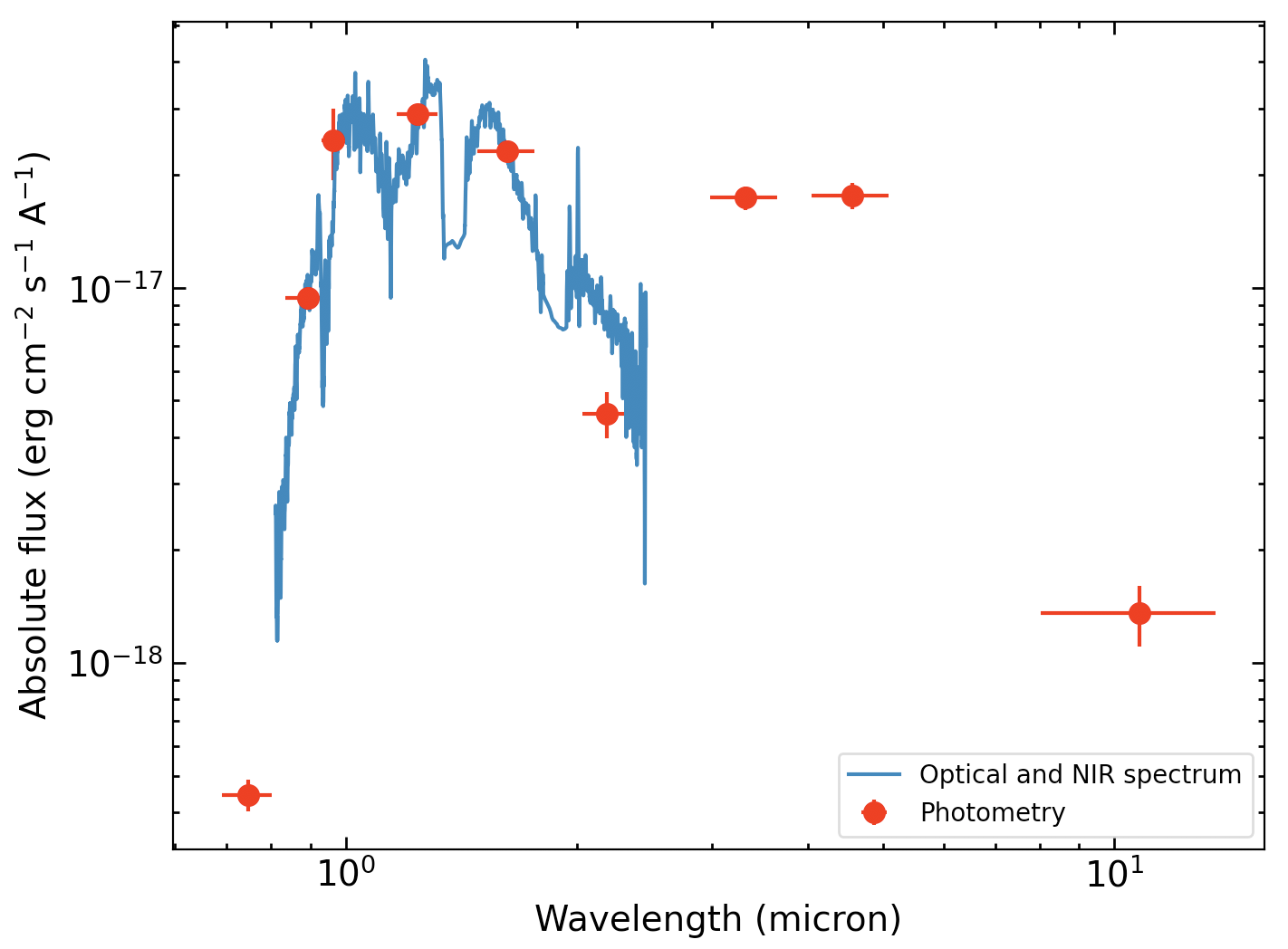}
   \caption{Combined photometric and spectroscopic spectral energy distribution of WISE1810. All OSIRIS, EMIR and \citet{schneider20a} spectra are smoothed for the clarity of the figure. Photometric fluxes (red dots) correspond to, from blue to red wavelengths, the $i$, $z$, $y$, $J$, $H$, $K$, $W1$, $W2$, and $W3$ bands. Horizontal error bars stand for the effective width of the filters while vertical error bars account for the photometric uncertainties.
   }
   \label{fig_esdT:sed}
\end{figure}
%

%
%%%%%%%%%%%%%%%%%%%%%%%%%%%%%%%%%%%%%
%%%%% Table: Logs for WISE1810 %%%%%
%%%%%%%%%%%%%%%%%%%%%%%%%%%%%%%%%%%%%
%
\begin{table}
 \centering
 \caption[]{Summary of physical parameters of WISE1810.
 }
 \begin{tabular}{@{\hspace{0.0mm}}l @{\hspace{2.0mm}}l @{\hspace{2mm}}l@{\hspace{0mm}}}
 \hline
 \hline
Name                  & Parameter                 &  Value  \cr
 \hline  
Parallax              & $\omega$                  & 112.5$^{+8.1}_{-8.0}$ mas \cr
Distance              & d                         & 8.9$^{+0.7}_{-0.6}$ pc \cr
Proper motion in RA   & $\mu_\alpha$cos$\delta$   & $-$1027.0$\pm$3.5 mas\,yr$^{-1}$ \cr
Proper motion in dec  & $\mu_\delta$              & $-$246.4$\pm$3.6 mas\,yr$^{-1}$ \cr
Heliocentric velocity & $v_{\rm h}$               & 45.6$\pm$3.5 km\,s$^{-1}$ \cr
Tangential velocity   & $v_{\rm t}$               & 44.5$\pm$3.6 km\,s$^{-1}$ \cr
Galactic velocity     & $U$                       & $-$36.9$\pm$2.9 km\,s$^{-1}$ \cr
Galactic velocity     & $V$                       & $-$44.5$\pm$1.8 km\,s$^{-1}$ \cr
Galactic velocity     & $W$                       & $-$29.1$\pm$2.7 km\,s$^{-1}$ \cr
Luminosity            & log\,($L/L_{\odot}$)      & $-$5.78$\pm$0.11 dex \cr 
Bolometric magnitude  & $M_{\rm bol}$             & 19.850$^{+0.082}_{-0.074}$ mag \cr
Effective temperature & $T_{\rm eff}$             & 800$\pm$100 K   \cr
Gravity               & log\,$g$                  & 5.0$\pm$0.25 dex (cm\,s$^{-2}$)\cr
Metallicity           & [Fe/H]                    & $-1.5 \pm 0.5$ dex  \cr
Radius                & $R$                       & 0.067$^{+0.032}_{-0.020}$ R$_{\odot}$ \cr
Mass                  & $M$                       & 17$^{+56}_{-12}$ M$_{\rm jup}$ \cr
\hline
 \label{tab_esdT:tab_properties}
 \end{tabular}
\end{table}
%

%
%%%%%%%%%%%%%%%%%%%%%%%%%%%%%%%%%%%%%%%%%%%%
%%%%% Figure: colour-luminosity
%%%%%%%%%%%%%%%%%%%%%%%%%%%%%%%%%%%%%%%%%%%%
% 
%
\begin{figure*}
   \centering
   \includegraphics[width=\linewidth, angle=0]{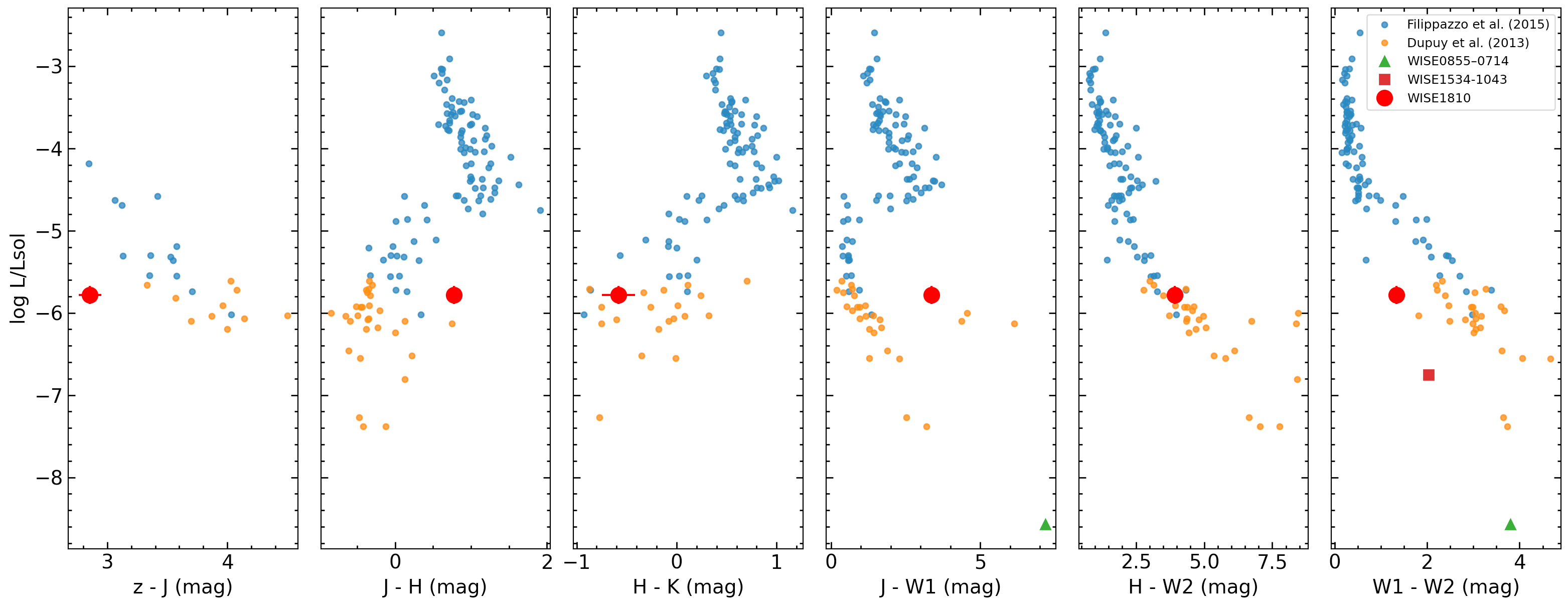}
   \caption{Colour--luminosity diagrams of solar-metallicity L, T, and Y dwarfs. L0--T7 dwarfs are plotted as blue dots while T8--Y1 dwarfs are shown as orange dots. WISE1810 is plotted as a red dot. In the right panel, and for comparison purposes, two objects: WISEA\,J153429.75$-$104303.3, the first Y subdwarf candidate \citep[known as ``the Accident'';][]{kirkpatrick21b}, and WISE\,J085510.83$-$071442.5, the nearest brown dwarf to the Sun at 2.2 pc \citep{luhman14b}.
   }
   \label{fig_esdT:colourlum}
\end{figure*}
\subsection{Bolometric luminosity}
\label{esdT:properties_Lbol}
The spectroscopic spectral energy distribution (SED) of WISE1810 was built using the optical OSIRIS spectrum, the near-infrared EMIR spectrum, and the $H$- and $K$-band spectrum by \citet{schneider20a}. It was flux calibrated using the $J$ and $H$ photometry (Table~\ref{tab:phot}). The spectroscopic SED was extended towards redder wavelengths by including the $W1-W3$ photometry given in Table~\ref{tab:phot}. We employed the filter profiles and zero point fluxes provided in the Spanish Virtual Observatory filters database \citep{rodrigo20} and the parallax determined here. The WISE1810 photometric and spectroscopic SED covering 0.6 through 16 $\mu$m is illustrated in Fig.\ \ref{fig_esdT:sed}. All fluxes in the figure are in the absolute scale, that is, they are corrected for the object's distance. There is a good agreement between the spectroscopic and photometric fluxes in all $zyJH$ bands; only the $K$ band is discrepant: the photometric determination indicates that WISE1810 has less flux at 2.2 $\mu$m than the fluxes provided by the observed spectra.

The bolometric luminosity of WISE1810 was derived by integrating the observed SED (Figure~\ref{fig_esdT:sed}); we obtained log\,$L/L_{\rm sol} = -5.78 \pm 0.11$ dex and $M_{\rm bol} = 19.850\,^{+0.082}_{-0.074}$ mag (Table \ref{tab_esdT:tab_properties}), where the error bars account for the photometric and parallax uncertainties. We neglected the fluxes below 0.6 $\mu$m and beyond 16 $\mu$m because they are likely very small and their contribution to the bolometric luminosity is presumably negligible. As a comparison, the bolometric luminosities of Luhman\,16A (L6--L7.5) and Luhman\,16B (T0$\pm$1) are log\,$L/L_{\rm sol} =  -4.66 \pm 0.08$ dex and $-4.68\pm0.13$ dex for effective temperatures around 1300\,K at a distance of 2.00$\pm$0.15 pc \citep{luhman13a,lodieu15b}, while the luminosity of WISE\,J085510.83–071442.5 \citep{luhman14b} at 2.2 pc is log\,$L/L_{\rm sol} = -8.57$ dex \citep{zapatero16}. The range of bolometric magnitudes and luminosities for the coolest brown dwarfs with spectral types between T8 and Y0 and effective temperatures between 400 and 730\,K lie between 17.7 and 20.9 mag and $-$5.6 and $-$6.6 dex, respectively \citep*{dupuy13b}. These intervals bracket our values for WISE1810\@. Although WISE1810 was classified as a metal-poor early-T dwarf by \citet{schneider20a}, its luminosity is comparable to old solar-metallicity T8--T9 dwarfs. This is illustrated in Figure~\ref{fig_esdT:colourlum}, where bolometric luminosities of L-, T- and Y-type dwarfs, and WISE\,J085510.83–071442.5 are plotted against different optical to infrared colours. The literature luminosities and magnitudes were extracted from \citet{leggett12b}, \citet{kirkpatrick12,kirkpatrick21c}, \citet{dupuy13b}, \citet{leggett17}, and \citet{zhang21}. WISE1810 has the same $H-K$ and $H-W2$ colours and bolometric luminosity as T8--T9 solar-metallicity dwarfs. WISE1810
is bluer in $W1-W2$, pointing towards a low metallicity as discussed in \citet{schneider20a} and \citet{kirkpatrick21b} and it is redder in the $J-H$ and $J-W1$ colours. In addition to these peculiarities, we found that it is bluer in the $z-J$ colour, which may also be attributed to a low metallicity. Its $i-z$ colour is typical of field T dwarfs \citep[Table 1 in][]{skrzypek15}. The right panel of  Figure~\ref{fig_esdT:colourlum} also includes the metal-depleted object WISEA\,J153429.75$-$104303.3 \citep{meisner20a} with an estimated temperature typical of Y dwarfs and located at a trigonometric distance of 16.3 pc \citep{kirkpatrick21b}. We determined this object's luminosity following the same method as for WISE1810, with the exception that "the Accident" (nicknamed by \citealt{kirkpatrick21b}) has no spectrum: we integrated over its photometric SED and obtained log $L/L_{\rm sol}$ = $-6.57 ^{+0.15}_{-0.16}$ dex. This is likely a lower limit on the bolometric luminosity because the photometric SED does not cover $H$- and $K$-band fluxes. Both WISE1810 and "the Accident (labeled WISE1534$-$1043 in the diagram) are subluminous in the luminosity versus $W1-W2$ diagram and appear to follow a sequence parallel to that of the "solar metallicity" dwarfs.

%
%%%%%%%%%%%%%%%%%%%%%%%%%%%%%%%%%%%%%%%%%%%%
%%%%% Figure: Spectra and models
%%%%%%%%%%%%%%%%%%%%%%%%%%%%%%%%%%%%%%%%%%%%
% 
%
\begin{figure*}
   \centering
   \includegraphics[width=\linewidth, angle=0]{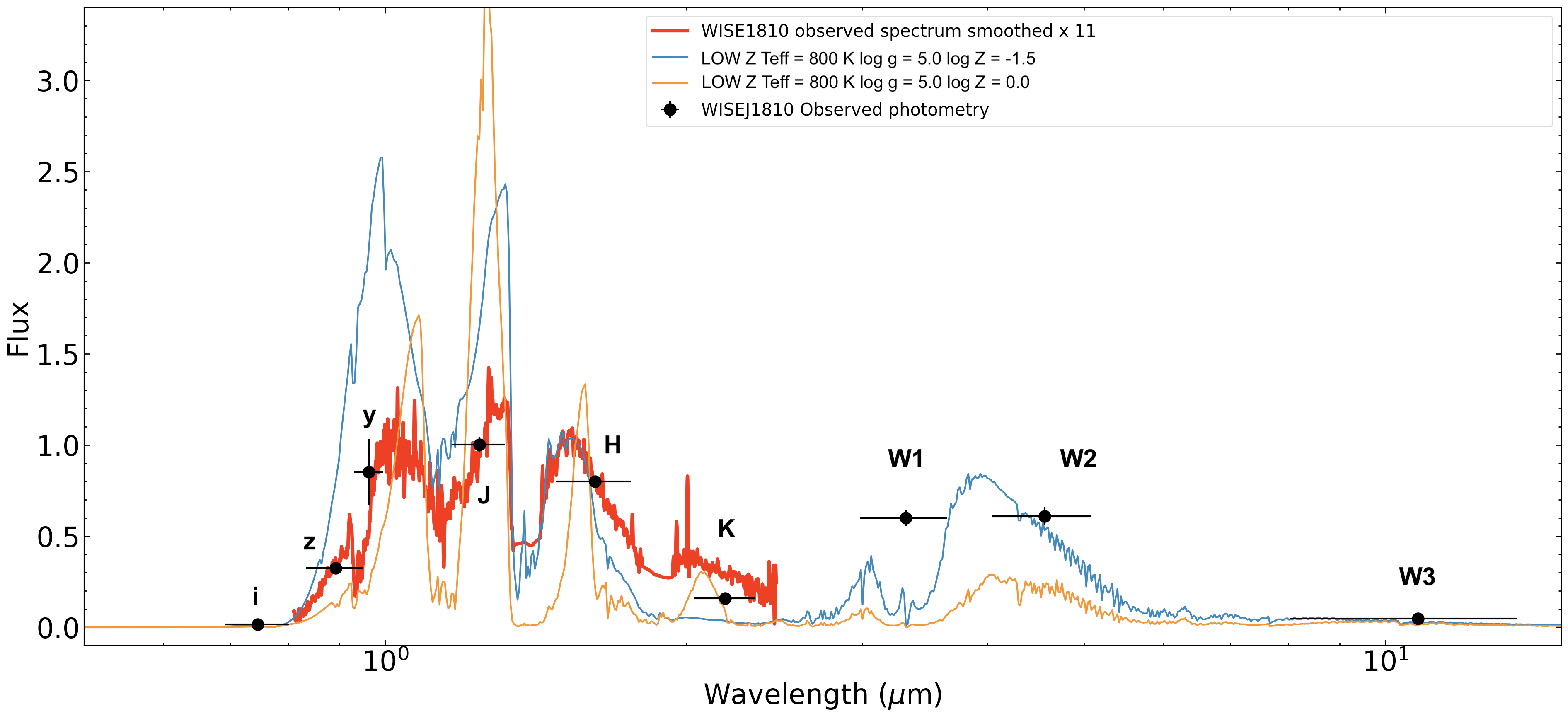}
   \caption{Comparison of the observed photometry (black dots) and spectrum (red line) of WISE1810 to LOWZ models \citep{meisner21}. Solar (blue) and low-metallicity [Fe/H] = $-1.5$ (orange) models are computed for $T_{\rm eff}$ = 800 K, log\,$g$ = 5.0 dex and a solar C/O ratio. The blue wing of the $H$ band and the $W2$ photometry are reasonably replicated by the metal-depleted theoretical spectrum. 
   All data are normalised between 1.5 and 1.6 $\mu$m. The wavelength axis is in the logarithmic scale.
   }
   \label{fig_esdT:spec_obs_model}
\end{figure*}
\subsection{Model fit of the spectrum of WISE1810}
\label{esdT:properties_model}

\begin{figure}
   \centering
   \includegraphics[width=\linewidth, angle=0]{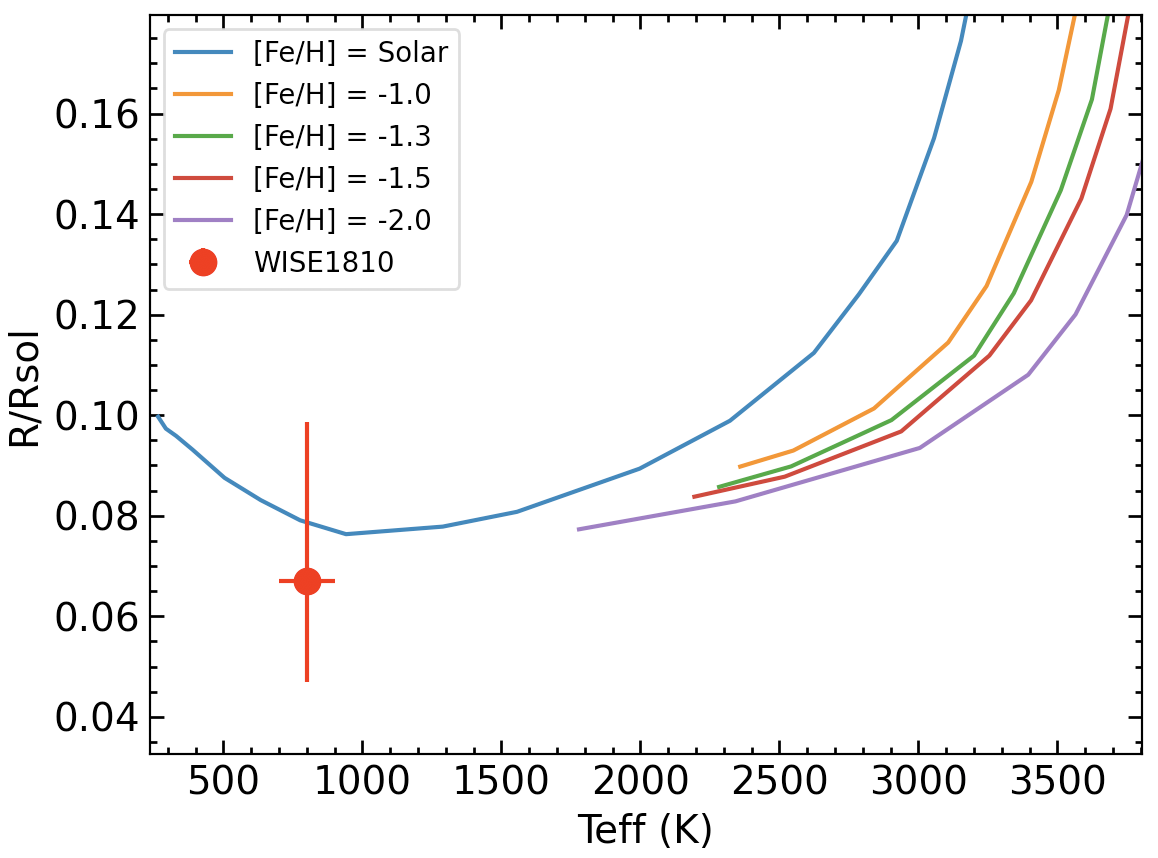}
   \caption{Dwarf radius vs temperature diagram. Models of different metallicities computed for an age of 10 Gyr (solid lines) are from \citet{baraffe97}. The metal depleted tracks cover all stellar masses down to the substellar boundary at 0.083 M$_\odot$ while the solar metallicity model extends into the brown dwarf regime. WISE1810 is depicted with the red dot.
   }
   \label{fig_esdT:teff_radius}
\end{figure}

We compared the WISE1810 SED to the recent LOWZ models presented in \citet{meisner21}. The LOWZ models were built to be useful for studies of brown dwarfs and were computed at a low spectral resolution for a wide range of metallicities (between [Fe/H] = $+$1.0 and $-$ 2.0 dex), effective temperatures ($T_{\rm eff}$ between 500 and 1600 K), and surface gravities (log\,$g$ between 3.5 and 5.5 in units of cm\,s$^{-2}$) with steps of 0.5 dex, 50--100 K, and 0.5 dex, respectively \citep{meisner21}. None of the LOWZ theoretical spectra is able to reproduce the entire observed SED of WISE1810. However, a few qualitative inferences can be made:
\begin{itemize}
    \item The shape of the observed $H$-band spectrum (particularly the blue wing) is better reproduced by the LOWZ models with high gravity (log\,$g$ $\approx 5.00 \pm 0.25$ dex) and low metallicity ([Fe/H] $\approx -1.5 \pm 0.25$ dex, see Fig.\ \ref{fig_esdT:spec_obs_model}). This could be explained by the less intense absorption of ammonia at these wavelengths.
    \item The flux at 4.5 $\mu$m ($W2$ band) is better reproduced by LOWZ models with $T_{\rm eff} \approx 800$ K (for log\,$g$ = 5 dex and [Fe/H] = $-$1.5 dex and a spectral normalisation at the $H$ band, see Fig.\ \ref{fig_esdT:spec_obs_model}). A similar criterion has been used to estimate the temperature of WISE\,J085510.83-071442.5 \citep{luhman14b,luhman14d}. Higher and lower temperatures yield smaller and larger predicted fluxes that are not compatible with the observations.
    \item The profile (shape and intensity) of the strong and broad water vapour absorption at $\sim$1.1 $\mu$m is not well replicated by the theory; however, metal-depleted models predict that this particular signature is present only at temperatures typically below 900--1000 K. It intensifies at lower metallicity. This result agrees with the temperature inferred from the flux emission at 4.5 $\mu$m.
    \item Models predict too strong collision-induced H$_2$ absorption affecting the 1--2.5 $\mu$m region and stronger fluxes between 0.8 and 1.3 $\mu$m. 
    \item WISE1810 is a water vapour dwarf; except for water vapour absorption and H$_2$ collision induced absorption, no other feature is visible in the SED. WISE1810 has no CO or CH$_4$ absorption at 2.2--2.5 $\mu$m and the theoretical flux at 3.6 $\mu$m (where the strongest CH$_4$ absorption lies) is clearly depleted as compared to the strong detection of WISE1810 in the $W1$ filter. This clearly contrasts with the previously inferred low temperatures at which methane absorption is expected (with the exception of the lowest metallicity). This suggests that WISE1810 has a C-deficient, metal-depleted atmosphere or alternatively, an oxygen-enhanced atmosphere. Unfortunately, the LOWZ models were computed only for C/O ratios of 0.1, solar, and 0.85.  
\end{itemize}
Despite the tension between the theory and the observations, and based on the LOWZ atmospheric models, we conclude that WISE1810 is likely a metal-depleted dwarf with a high-gravity atmosphere, [Fe/H] $\approx -1.5$ dex, and $T_{\rm eff} \le 1000$ K (possibly around 800 K, Table \ref{tab_esdT:tab_properties}). For comparison, \citet{schneider20a} inferred an effective temperature of 1300$\pm$100\,K\@. With such a low temperature, WISE1810 is very likely a substellar object ($M < 0.08$ M$_\odot$) and not a low-mass star. The least massive stars ($\approx$ 0.08 M$_\odot$) with [Fe/H] = $-1.5$ dex have predicted temperatures of about 2000 K and luminosities $\sim$100 times higher than WISE1810 at the age of 10 Gyr \citep{baraffe97}, as illustrated in Fig.\ \ref{fig_esdT:teff_radius}. 
This figure shows that all very small stars have similar radii of about 0.09$\pm$0.02 R$_{\odot}$ and that intermediate-mass to massive brown dwarfs have related sizes (at least for the solar metallicity case). Therefore, for a given age, luminosity mainly depends on effective temperature.
At a given mass (e.g.\ 0.085 M$_{\odot}$) \cite{baraffe97} predict bolometric luminosities of log\,$L/L_{\rm sol}$ = $-$3.49, $-$3.56, and $-$3.75 dex for objects with [Fe/H] = $-$1.0, $-$1.5, and $-$2.0 dex, respectively. It is important that these results are fully model dependent, that is, they must be revised when well-tested low-metallicity model atmospheres become available.

%
%%%%%%%%%%%%%%%%%%%%%%%%%%%%%%%%%%%%%%%%%%%%
%%%%% Figure: Teff vs Radius
%%%%%%%%%%%%%%%%%%%%%%%%%%%%%%%%%%%%%%%%%%%%
% 
%

\subsection{Mass and radius}

Using the luminosity--radius--temperature relation of the Stefan-Boltzmann law ($L = 4 \pi R^{2} \sigma T^{4}_{\rm eff}$), and adopting the WISE1810 luminosity given in section~\ref{esdT:properties_Lbol} and $T_{\rm eff} = 800 \pm 100$ K, we derived the radius $R = 0.067 ^{+0.032}_{-0.020}$ R$_\odot$ (or 0.67 times the radius of Jupiter) for WISE1810 (Table \ref{tab_esdT:tab_properties}). 
The large error bar is due to the large uncertainty in the temperature determination (temperature goes to the fourth power in the relation). This value is only slightly below the minimum size (0.076 R$_\odot$) predicted by theoretical evolutionary models for solar metallicity brown dwarfs \citep{baraffe97}, see Figure~\ref{fig_esdT:teff_radius}. To the best of our knowledge, there are no substellar evolutionary tracks computed for low chemical abundances in the literature. Our size determination for WISE1810 indicates that metal-depleted brown dwarfs also have a size similar to (or slightly smaller than) their solar-abundance counterparts, and consequently they are close in size to Jupiter.

Using Newton's law of universal gravitation ($g = $, solar surface gravity log\,$g_{\rm sol}$ = 4.44 (cm\,s$^{-2}$), solar $T_{\rm eff}$ = 5777 K, and the surface gravity (log\,$g$), $T_{\rm eff}$, and bolometric luminosity inferred for WISE1810, we derived a mass of $0.016^{+0.048}_{-0.012}$ M$_\odot$ (or $17^{+56}_{-12}$ M$_{\rm Jup}$) according to the following equation:
\begin{equation}
    {\rm log}\,(M/M_{\rm sol}) = {\rm log}\,g + {\rm log}\,(L/L_{\rm sol}) + 4\,{\rm log}\,(\frac{5777}{T_{\rm eff}}) - {\rm log}\,g_{\rm sol}
\end{equation}
The mass uncertainty accounts for the error bars of all the WISE1810 parameters (Table \ref{tab_esdT:tab_properties}). Our mass determination is below the original mass estimate of 0.075--0.080 M$_{\odot}$ \citep{schneider20a}. At the 1$\sigma$ confidence level, WISE1810 has a mass below the star--brown dwarf borderline for low metallicity \citep{baraffe97}, thus making it a genuine metal-depleted brown dwarf. However, the mass uncertainty is rather large and we cannot conclude whether it is a massive or a low-mass brown dwarf near the brown dwarf--planet boundary. An improvement of the model atmospheres is needed to determine more precise surface gravity and temperature that would lead to a more accurate mass calculation.

%
%%%%%%%%%%%%%%%%%%%%%%%%%%%%%%%%%%%%%%%%%%%%
%%%%% Figure: UVW space motion
%%%%%%%%%%%%%%%%%%%%%%%%%%%%%%%%%%%%%%%%%%%%
% 
%
\begin{figure*}
   \centering
   \includegraphics[width=\linewidth, angle=0]{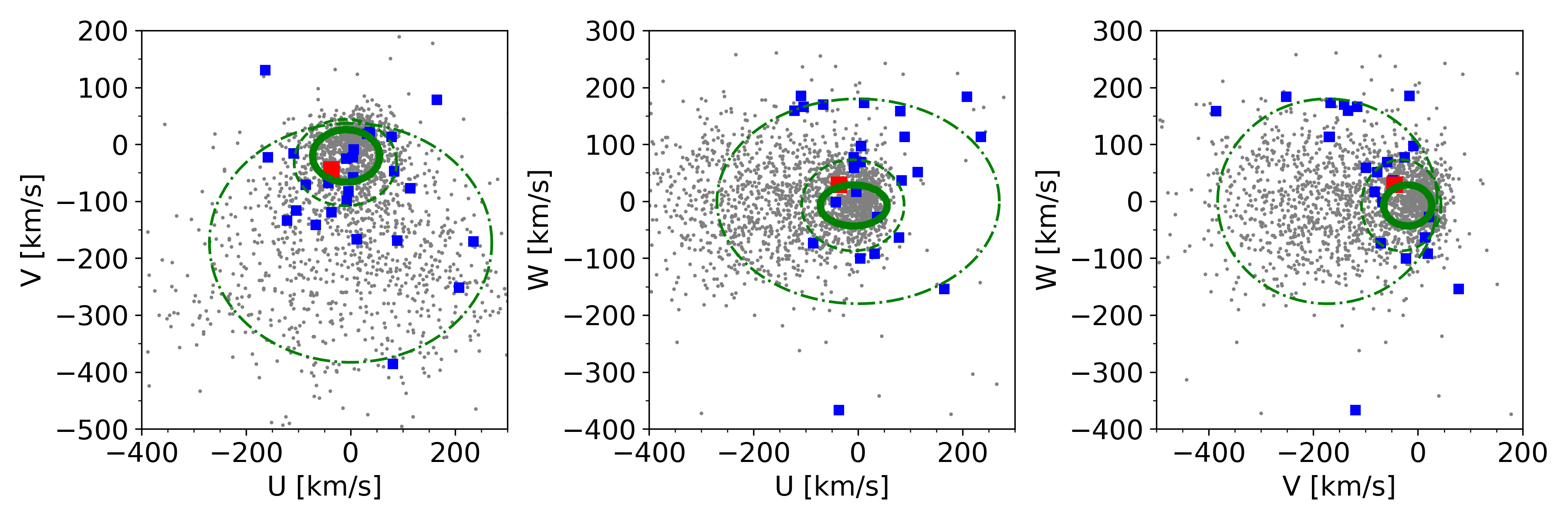}
   \caption{Space motion of WISE1810 (red square) compared with the galactic velocities of L subdwarfs (blue squares) and M subdwarfs from the sample of \citet{savcheva14} with $\zeta$\,$\leq$\,0.825 (grey dots). The 2$\sigma$ dispersion of the disc, thick disc, and halo populations are drawn as the dash-dotted, dashed, and solid circles, respectively \citep{fuchs09a,savcheva14}.
   }
   \label{fig_esdT:space_motion}
\end{figure*}
\subsection{Galactic velocity}
\label{esdT:properties_UVW}

We used two methods to attempt a measurement of the heliocentric radial velocity ($v_{\rm h}$) for WISE1810. 
First, we cross-correlated the spectral region around the
Cs{\small{I}} resonance line at 894.3 nm, where the S/N of the optical spectrum is high, against a template dwarf with a known velocity: DENIS\,J12281523$-$1547342 
\citep{delfosse97,martin97a,faherty12,dupuy12}, for which we had collected a high-quality spectrum in April 2019 with the same GTC OSIRIS instrumental configuration \citep{martin18a}.
We derived $v_{\rm h}$\,=\,$-$48.6$\pm$3.6 km/s for WISE1810 measured by fitting the cross-correlation function with a Gaussian profile using the IRAF task {\tt{fxcor}}.

Second, we cross-correlated the spectral region around the water band between 1.10 and 1.16 $\mu$m, the molecular feature dominating in the spectral range covered by the GTC EMIR spectrum of WISE1810, against the spectroscopic standard observed at similar airmass to WISE1810 to perform the correction for telluric absorption. We used the WISE1810 spectrum uncorrected for tellurics because the telluric-corrected spectrum was too noisy. 
We did not see any peaks in the correlation function that could be attributed to the telluric contribution and thus we consider that the dominant source of steam opacity in this spectral region is the atmosphere of WISE1810 itself. The radial velocity value and its error bar were measured by fitting the cross-correlation function with a Gaussian profile using the IRAF task {\tt{fxcor}}.
We obtained $v_{\rm h}$\,=\,$-$45.6$\pm$3.5 km/s for WISE1810 (Table \ref{tab_esdT:tab_properties}). Both independent measurements of radial velocity using the EMIR and OSIRIS spectra agree quite well within the uncertainties, suggesting a mean value of $v_{\rm h}$\,=\,$-$47.1$\pm$2.6 km/s. Nevertheless, we do not attach a high significance to this value because both methods are hampered by the presence of telluric lines, which are very difficult to properly take into account at the modest spectral resolution of our spectra. A reliable measurement of the $v_{\rm h}$ for WISE1810 requires higher resolution spectroscopy over a spectral region with sufficient Doppler information, which may be difficult to find for such a low-metallicity, relatively featureless object.

With our new trigonometric distance and tentative radial velocity, we infer a galactic motion of ($U$, $V$, $W$)\,=\,($-36.9 \pm 2.5, -44.5 \pm 1.8, 29.1 \pm 2.7$) km\,s$^{-1}$ for WISE1810 (Table \ref{tab_esdT:tab_properties}) using the formulation of \citet*{johnson87}. The velocity $U$ is positive towards the Galactic centre, $V$ is positive in the direction of Galactic rotation, and $W$ is positive towards the north Galactic pole. 
On the one hand, the kinematics of WISE1810 is not consistent with membership to any of the young stellar streams in the solar neighbourhood as was expected from its likely high gravity and low-metallicity atmosphere. On the other hand, none of the Galactic velocity components is indicative of the membership of WISE1810 to the halo of the Galaxy (Fig.\ref{fig_esdT:space_motion}): all velocities are low, particularly the $V$ and $W$ components, compared to the typical velocities of hundreds of km\,s$^{-1}$ resulting from the dynamical heating of the old halo stars. The WISE1810 kinematics are at the borderline between the Galactic thin and thick disc populations (Fig.\ \ref{fig_esdT:space_motion}), see also \citep[Figure 17 of][]{zhang17a}. 

Given the high uncertainty in our determination of the heliocentric velocity, we also explored the kinematic properties of WISE1810 via its tangential velocity, $v_{\rm t} = 44.5 \pm 3.6$ km\,s$^{-1}$, which is directly computed from the trigonometric parallax and proper motion. We compared it to the distribution of the tangential velocities of the 20 pc sample of L-, T-, and Y-type dwarfs \citep{kirkpatrick21c}. Only $\sim$28\,\% of the L, T, and Y dwarfs in the solar vicinity show a tangential velocity greater than that of WISE1810. Despite the marked spectroscopic and photometric differences between WISE1810 and the great majority of the L, T, and Y dwarfs within 20 pc, they all share  similar kinematics (i.e.\ WISE1810 does not stand out for its kinematics), which may indicate that WISE1810 is not a halo member.

In Fig.\ \ref{fig_esdT:space_motion}, we compared the space motion of WISE1810 with known L subdwarfs \citep[]{zhang18b} and M subdwarfs with spectral types from the Sloan Digital Sky Survey \citep{savcheva14} and the $\zeta_{TiO/CaH}$ parameter defined as (1-TiO5)/(1-TiO5$_{Z_{\odot}}$) below 0.825 \citep{lepine07c}\footnote{The TiO5, CaH1, CaH2, and CaH3 indices were originally defined in \citet{gizis97a} and later revised by \citet{lepine03a}}. On the one hand, we observe that WISE1810 has space motions similar to a few L subdwarfs with velocities similar to the disc population. On the other hand, many M subdwarfs exhibit kinematics consistent with the disc population. Although WISE1810 does not seem to be a member of the halo, this is not necessarily inconsistent with its metal-poor nature.

\subsection{Density of metal-poor brown dwarfs}
\label{esdT:properties_density}

As discussed in the previous section, WISE1810 is very likely a substellar object 
due to its low luminosity and cool effective temperature.
Our updated distance of 8.9 pc makes WISE1810 a new addition to the 10 pc sample 
\citep{henry06,henry18,reyle18,reyle21}.
The last authors have compiled 4 A, 8 F, 18 G, 38 K, 249 M, 21 L, 45 T, and 19 Y dwarfs 
within 10 pc; their number of ultra-cool dwarfs is slightly larger than those
 compiled in \citet[17 L, 41 T, 18 Y;][]{kirkpatrick21c}.
The ratio of known ultra-cool dwarfs with spectral type L and later (85)
to that of warmer objects with spectral type M and earlier (317) is 0.27\@. 
For comparison, this ratio was only 0.04 in 2010 according to the RECONS database. 
Our parallax measurement for WISE1810 reinforces the idea that the census of 
substellar objects is still very incomplete even in our immediate neighbourhood.

Among the L, T, Y dwarfs in the 10 pc sample presented by \citet{,kirkpatrick21c}, 
three sources are classified as L and T subdwarfs, that is, dwarfs with low metallicity (although the exact metallicity values remain unknown): the two components of the
pair formed by SDSS\,J141624.08$+$134826.7 (sdL7) and ULAS\,J141623.94$+$134836.3 
\citep[sdT7;][]{burningham10a,schmidt10a,scholz10a,bowler10a}
and CWISE\,J105512.11$+$544328.3 \citep[sdT8;][]{kirkpatrick21c}.
However, we should note that the retrieval of the spectra of the primary and secondary of J1416$+$1348 indicate a metallicity between $-$0.5 dex and solar.
In addition, four objects are classified as peculiar possibly with low metallicity (sd): 
2MASS\,J0729000$-$395404 \citep[$\pi$\,=\,126.3 mas; T8pec;][]{looper07,faherty12} with a reduced flux in the $H$ and $K$ bands; 
2MASSI\,J0937347$+$293142 \citep[$\pi$\,=\,162.8 mas; T6pec;][]{burgasser02,burgasser06a,vrba04}
peculiar in the $K$ band,
2MASS\,J0939355$-$2448279A \citep[$\pi$\,=\,187.3 mas; T8;][]{tinney05,burgasser06a,burgasser08d} with
a metallicity between $-$0.3 dex and solar and
WISEPC\,J232519.5$-$410534.9  \citep[$\pi$\,=\,108.4 mas; T9pec;][]{kirkpatrick11,tinney14,kirkpatrick19}
highlighted as possibly metal poor.
Overall, the ratio of moderately low-metallicity late-type dwarfs over the total number of L, T, and Y dwarfs lies in the interval 1.3--9.2\,\%~(1--7/76). 
Adding WISE1810 to the census of late-type candidates within 10 pc, this ratio of low-metallicity ultra-cool dwarfs over the population of ultra-cool dwarfs would increase to 2.6--10.4\,\%~(2--8/77).
This frequency of metal-poor brown dwarfs is much higher than the disc-to-halo ratio of 
1-to-200 derived from main-sequence K subdwarfs or evolved red giants \citep{juric08a},
the contribution of FGK subdwarfs identified in photographic plates \citep[0.2\%;][]{digby03} 
and M subdwarfs selected photometrically in large-scale surveys \citep[0.65\%][]{covey08a}, 
whose metallicity is below $-$0.5 dex \citep{lodieu19c}. 

In the catalogue of M dwarfs targeted by CARMENES \citep{quirrenbach20}, \citet{passegger18} derived 
physical parameters for 300 targets whose metallicities range between $-$0.43 dex and $+$0.34 dex. 
Similarly, \citet{marfil21} investigated the temperature, gravity, and metallicity of 343 M dwarfs 
part of the CARMENES survey. None of the M dwarfs has metallicities below $-$0.7 dex and only 
12 have metallicities below $-$0.5 dex, yielding a strict upper limit of 3.5\% of slightly metal-poor M dwarfs 
in the CARMENES sample. However, none of these M dwarfs has metallicities below Fe/H\,=\,$-$1.0 dex 
and none is classified as a subdwarf from low-resolution optical spectroscopy \citep{alonso-floriano15a}, 
placing an upper limit of 0.3\% of M subdwarfs in the CARMENES sample. Similarly, only one source
within 20 pc (Gl\,809) is classified as a subdwarf among the 1564 bright ($J$\,$\leq$9 mag)
M dwarfs in the catalogue of \citet{lepine13a}, resulting in a ratio of 0.064\% broadly
consistent with the proportion of late-type subdwarfs to solar-metallicity stars from a photometric sample of $\sim$100 sources with spectroscopic follow-up \citep[$\sim$0.02\%][]{lodieu12b,lodieu17a}.
The most recent study of the 20 pc sample with $Gaia$ \citep{reyle18} suggests that only 4 sources
out of 1544 with radial velocities have (V,W) galactocentric velocities consistent with the halo 
as defined by \citet{reddy06} and \citet{zhang17a}, yielding a fraction of metal-poor stars of 0.26\%. 
Considering stars within 100 pc, this fraction increases significantly with about 2.3\% of objects with 
astrometry typical of halo stars. In spite of the large uncertainties due to the inhomogenous surveys
targeting metal-poor stars, we find a frequency in the range 0.064--0.65\% with a strict upper limit of 3.5\%.
To summarise, although there is an overlap in the fraction of low-metallicity ultra-cool dwarfs and stars, current statistics indicate that the former population is likely more abundant in relative terms.
More metal-poor brown dwarfs are needed to further constrain their frequency.

%
%%%%%%%%%%%%%%%%%%%%%%%%%%%%%%%%%%%%%%%%%%%%
%%%%% Conclusions %%%%%
%%%%%%%%%%%%%%%%%%%%%%%%%%%%%%%%%%%%%%%%%%%%
%
\section{Conclusions}
\label{esdT:conclusions}

The main focus of this paper is WISE1810, the closest extreme ultra-cool subdwarf
to the Sun straddling the hydrogen-burning limit. We present for the first time
a low-resolution optical spectrum as well as an improved infrared spectrum. 
We inferred a ground-based trigonometric parallax of 112.5 mas, placing WISE1810 
at 8.9$^{+0.7}_{-0.6}$ pc. Hence, WISE1810 represents a new important addition to the 10 pc sample.
We inferred a luminosity of $-$5.78 dex, typical of solar-metallicity late-T dwarfs 
in the solar neighbourhood, implying a very cool effective temperature for WISE1810\@.
No currently available theoretical atmosphere model reproduces the optical-to-infrared spectrum of WISE1810\@.
Finally, we conclude that the density of metal-poor brown dwarfs in the solar neighbourhood
might be higher than expected due to the revised distance of WISE1810 and the recent 
Y-type `accident' reported by \citet{kirkpatrick21b}. 

%
%%%%%%%%%%%%%%%%%%%%%%%%%%%%%%%%%%%%%%%%%%%%
%%%%% Acknowledgements %%%%%
%%%%%%%%%%%%%%%%%%%%%%%%%%%%%%%%%%%%%%%%%%%%
%
\begin{acknowledgements}
NL and ELM acknowledge  support  from  the Agencia  Estatal  de  Investigaci\'on  del  Ministerio  de  Ciencia  e Innovaci\'on (AEI-MCINN) under grant PID2019-109522GB-C53\@. 
MRZO acknowledges  support  from  the Agencia  Estatal  de  Investigaci\'on  del  Ministerio  de  Ciencia  e Innovaci\'on (AEI-MCINN) under grant PID2019-109522GB-C51\@.
BG acknowledge support from the UK Science and Technology Facilities Council (STFC) via the Consolidated Grant ST/R000905/1\@.
% Thanks to other colleagues
NL warmly thank Adam Schneider for sharing the infrared spectra published
in \citet{schneider20a} and Adam Burgasser for sending his T dwarf optical templates. We appreciate the scientific discussions on the nature of the object with our colleague V\'ictor J$.$ S$.$ B\'ejar.
% NOT
The data presented here were obtained with NOT/ALFOSC, which is provided by the Instituto de Astrofisica de Andalucia (IAA) under a joint agreement with the University of Copenhagen and NOTSA\@.
% GTC
Based on observations made with the Gran Telescopio Canarias (GTC), in the Spanish Observatorio del Roque de los Muchachos of the Instituto de Astrofísica de Canarias, under Director’s Discretionary Time programme number GTC03-20BDDT and GTC05-20BDDT and normal programme GTCMULTIPLE3A-21A\@.
% Calar Alto
Based on observations collected at the Centro Astronómico Hispano-Alemán (CAHA) at Calar Alto, operated jointly by Junta de Andalucía and Consejo Superior de Investigaciones Científicas (IAA-CSIC) under programmes H20-3.5-020 and F21-3.5-010\@.
% CDS+ADS
This research has made use of the Simbad and Vizier databases, operated
at the centre de Donn\'ees Astronomiques de Strasbourg (CDS), and
of NASA's Astrophysics Data System Bibliographic Services (ADS).
% UKIDSS
The UKIDSS project is defined in \citet{lawrence07}. UKIDSS uses the UKIRT Wide Field Camera
\citet[WFCAM;][]{casali07}. The photometric system is described in \citet{hewett06}, and the
calibration is described in \citet{hodgkin09}. The pipeline processing and science archive are
described in \citet{irwin04} and \citet{hambly08}. \\
% ESO VVV
Based on data products from observations made with ESO Telescopes at the La Silla or Paranal Observatories under ESO programme ID 179.B-2002 \citep{minniti10,saito10a}. \\
% AllWISE
This publication makes use of data products from the Wide-field Infrared Survey Explorer, which
is a joint project of the University of California, Los Angeles, and the Jet Propulsion
Laboratory/California Institute of Technology, and NEOWISE, which is a project of the Jet
Propulsion Laboratory/California Institute of Technology. WISE and NEOWISE are funded
by the National Aeronautics and Space Administration. \\
%PanStarrs
The Pan-STARRS1 Surveys (PS1) and the PS1 public science archive have been made possible
through contributions by the Institute for Astronomy, the University of Hawaii, the Pan-STARRS
Project Office, the Max-Planck Society and its participating institutes, the Max Planck Institute
for Astronomy, Heidelberg and the Max Planck Institute for Extraterrestrial Physics, Garching,
The Johns Hopkins University, Durham University, the University of Edinburgh, the Queen's
University Belfast, the Harvard-Smithsonian Center for Astrophysics, the Las Cumbres Observatory
Global Telescope Network Incorporated, the National Central University of Taiwan, the Space
Telescope Science Institute, the National Aeronautics and Space Administration under
Grant No.\ NNX08AR22G issued through the Planetary Science Division of the NASA Science
Mission Directorate, the National Science Foundation Grant No. AST-1238877, the University
of Maryland, Eotvos Lorand University (ELTE), the Los Alamos National Laboratory, and the
Gordon and Betty Moore Foundation. \\
%IRSA
This research has made use of the NASA/IPAC Infrared Science Archive, which is funded by the National Aeronautics and Space Administration and operated by the California Institute of Technology.
\end{acknowledgements}
%

%
%%%%%%%%%%%%%%%%%%%%%%%%%%%%%%%%%%%%%%%%%%%%
%%%%% Bibliography %%%%%
%%%%%%%%%%%%%%%%%%%%%%%%%%%%%%%%%%%%%%%%%%%%
%
\bibliographystyle{aa} % style aa.bst
\bibliography{mnemonic,biblio_old} % your references Yourfile.bib

\begin{thebibliography}{118}
\expandafter\ifx\csname natexlab\endcsname\relax\def\natexlab#1{#1}\fi

\bibitem[{{Allard} {et~al.}(2012){Allard}, {Homeier}, \& {Freytag}}]{allard12}
{Allard}, F., {Homeier}, D., \& {Freytag}, B. 2012, Royal Society of London
  Philosophical Transactions Series A, 370, 2765

\bibitem[{{Alonso-Floriano} {et~al.}(2015){Alonso-Floriano}, {Morales},
  {Caballero}, {Montes}, {Klutsch}, {Mundt}, {Cort{\'e}s-Contreras}, {Ribas},
  {Reiners}, {Amado}, {Quirrenbach}, \& {Jeffers}}]{alonso-floriano15a}
{Alonso-Floriano}, F.~J., {Morales}, J.~C., {Caballero}, J.~A., {et~al.} 2015,
  A\&A, 577, A128

\bibitem[{{Baraffe} {et~al.}(1997){Baraffe}, {Chabrier}, {Allard}, \&
  {Hauschildt}}]{baraffe97}
{Baraffe}, I., {Chabrier}, G., {Allard}, F., \& {Hauschildt}, P.~H. 1997, A\&A,
  327, 1054

\bibitem[{{Bowler} {et~al.}(2010){Bowler}, {Liu}, \& {Dupuy}}]{bowler10a}
{Bowler}, B.~P., {Liu}, M.~C., \& {Dupuy}, T.~J. 2010, ApJ, 710, 45

\bibitem[{{Brooks} {et~al.}(2022){Brooks}, {Kirkpatrick}, {Caselden},
  {Schneider}, {Meisner}, {Faherty}, {Casewell}, {Kuchner}, {Kuchner}, \&
  {Backyard Worlds: Planet 9 Collaboration}}]{brooks22a}
{Brooks}, H., {Kirkpatrick}, J.~D., {Caselden}, D., {et~al.} 2022, AJ, 163, 47

\bibitem[{{Burgasser} {et~al.}(2006){Burgasser}, {Geballe}, {Leggett},
  {Kirkpatrick}, \& {Golimowski}}]{burgasser06a}
{Burgasser}, A.~J., {Geballe}, T.~R., {Leggett}, S.~K., {Kirkpatrick}, J.~D.,
  \& {Golimowski}, D.~A. 2006, ApJ, 637, 1067

\bibitem[{{Burgasser} {et~al.}(2002){Burgasser}, {Kirkpatrick}, {Brown},
  {Reid}, {Burrows}, {Liebert}, {Matthews}, {Gizis}, {Dahn}, {Monet}, {Cutri},
  \& {Skrutskie}}]{burgasser02}
{Burgasser}, A.~J., {Kirkpatrick}, J.~D., {Brown}, M.~E., {et~al.} 2002, ApJ,
  564, 421

\bibitem[{{Burgasser} {et~al.}(2003{\natexlab{a}}){Burgasser}, {Kirkpatrick},
  {Burrows}, {Liebert}, {Reid}, {Gizis}, {McGovern}, {Prato}, \&
  {McLean}}]{burgasser03b}
{Burgasser}, A.~J., {Kirkpatrick}, J.~D., {Burrows}, A., {et~al.}
  2003{\natexlab{a}}, ApJ, 592, 1186

\bibitem[{{Burgasser} {et~al.}(2000){Burgasser}, {Kirkpatrick}, {Cutri},
  {McCallon}, {Kopan}, {Gizis}, {Liebert}, {Reid}, {Brown}, {Monet}, {Dahn},
  {Beichman}, \& {Skrutskie}}]{burgasser00a}
{Burgasser}, A.~J., {Kirkpatrick}, J.~D., {Cutri}, R.~M., {et~al.} 2000, ApJL,
  531, L57

\bibitem[{{Burgasser} {et~al.}(2003{\natexlab{b}}){Burgasser}, {Kirkpatrick},
  {Liebert}, \& {Burrows}}]{burgasser03d}
{Burgasser}, A.~J., {Kirkpatrick}, J.~D., {Liebert}, J., \& {Burrows}, A.
  2003{\natexlab{b}}, ApJ, 594, 510

\bibitem[{{Burgasser} {et~al.}(2008){Burgasser}, {Tinney}, {Cushing}, {Saumon},
  {Marley}, {Bennett}, \& {Kirkpatrick}}]{burgasser08d}
{Burgasser}, A.~J., {Tinney}, C.~G., {Cushing}, M.~C., {et~al.} 2008, ApJL,
  689, L53

\bibitem[{{Burningham} {et~al.}(2010){Burningham}, {Leggett}, {Lucas},
  {Pinfield}, {Smart}, {Day-Jones}, \& {8 co-authors}}]{burningham10a}
{Burningham}, B., {Leggett}, S.~K., {Lucas}, P.~W., {et~al.} 2010, MNRAS, 404,
  1952

\bibitem[{{Burningham} {et~al.}(2014){Burningham}, {Smith}, {Cardoso}, {Lucas},
  {Burgasser}, {Jones}, \& {Smart}}]{burningham14}
{Burningham}, B., {Smith}, L., {Cardoso}, C.~V., {et~al.} 2014, MNRAS, 440, 359

\bibitem[{{Casali} {et~al.}(2007){Casali}, {Adamson}, {Alves de Oliveira},
  {Almaini}, {Burch}, {Chuter}, {Elliot}, \& {23 co-authors}}]{casali07}
{Casali}, M., {Adamson}, A., {Alves de Oliveira}, C., {et~al.} 2007, A\&A, 467,
  777

\bibitem[{{Cepa} {et~al.}(2000){Cepa}, {Aguiar}, {Escalera},
  {Gonzalez-Serrano}, {Joven-Alvarez}, {Peraza}, {Rasilla}, {Rodriguez-Ramos},
  {Gonzalez}, {Cobos Duenas}, {Sanchez}, {Tejada}, {Bland-Hawthorn},
  {Militello}, \& {Rosa}}]{cepa00}
{Cepa}, J., {Aguiar}, M., {Escalera}, V.~G., {et~al.} 2000, in Society of
  Photo-Optical Instrumentation Engineers (SPIE) Conference Series, Vol. 4008,
  Society of Photo-Optical Instrumentation Engineers (SPIE) Conference Series,
  ed. {M.~Iye \& A.~F.~Moorwood}, 623--631

\bibitem[{{Chambers} {et~al.}(2016){Chambers}, {Magnier}, {Metcalfe},
  {Flewelling}, {Huber}, {Waters}, {Denneau}, {Draper}, {Farrow}, {Finkbeiner},
  {Holmberg}, {Koppenhoefer}, {Price}, {Rest}, {Saglia}, {Schlafly}, {Smartt},
  {Sweeney}, {Wainscoat}, {Burgett}, {Chastel}, {Grav}, {Heasley}, {Hodapp},
  {Jedicke}, {Kaiser}, {Kudritzki}, {Luppino}, {Lupton}, {Monet}, {Morgan},
  {Onaka}, {Shiao}, {Stubbs}, {Tonry}, {White}, {Ba{\~n}ados}, {Bell},
  {Bender}, {Bernard}, {Boegner}, {Boffi}, {Botticella}, {Calamida},
  {Casertano}, {Chen}, {Chen}, {Cole}, {Deacon}, {Frenk}, {Fitzsimmons},
  {Gezari}, {Gibbs}, {Goessl}, {Goggia}, {Gourgue}, {Goldman}, {Grant},
  {Grebel}, {Hambly}, {Hasinger}, {Heavens}, {Heckman}, {Henderson}, {Henning},
  {Holman}, {Hopp}, {Ip}, {Isani}, {Jackson}, {Keyes}, {Koekemoer}, {Kotak},
  {Le}, {Liska}, {Long}, {Lucey}, {Liu}, {Martin}, {Masci}, {McLean}, {Mindel},
  {Misra}, {Morganson}, {Murphy}, {Obaika}, {Narayan}, {Nieto-Santisteban},
  {Norberg}, {Peacock}, {Pier}, {Postman}, {Primak}, {Rae}, {Rai}, {Riess},
  {Riffeser}, {Rix}, {R{\"o}ser}, {Russel}, {Rutz}, {Schilbach}, {Schultz},
  {Scolnic}, {Strolger}, {Szalay}, {Seitz}, {Small}, {Smith}, {Soderblom},
  {Taylor}, {Thomson}, {Taylor}, {Thakar}, {Thiel}, {Thilker}, {Unger},
  {Urata}, {Valenti}, {Wagner}, {Walder}, {Walter}, {Watters}, {Werner},
  {Wood-Vasey}, \& {Wyse}}]{chambers16}
{Chambers}, K.~C., {Magnier}, E.~A., {Metcalfe}, N., {et~al.} 2016, arXiv
  e-prints, arXiv:1612.05560

\bibitem[{{Covey} {et~al.}(2008){Covey}, {Hawley}, {Bochanski}, {West}, {Reid},
  {Golimowski}, {Davenport}, {Henry}, {Uomoto}, \& {Holtzman}}]{covey08a}
{Covey}, K.~R., {Hawley}, S.~L., {Bochanski}, J.~J., {et~al.} 2008, AJ, 136,
  1778

\bibitem[{{Cushing} {et~al.}(2011){Cushing}, {Kirkpatrick}, {Gelino},
  {Griffith}, {Skrutskie}, {Mainzer}, {Marsh}, {Beichman}, {Burgasser},
  {Prato}, {Simcoe}, {Marley}, {Saumon}, {Freedman}, {Eisenhardt}, \&
  {Wright}}]{cushing11}
{Cushing}, M.~C., {Kirkpatrick}, J.~D., {Gelino}, C.~R., {et~al.} 2011, ApJ,
  743, 50

\bibitem[{{Delfosse} {et~al.}(1997){Delfosse}, {Tinney}, {Forveille},
  {Epchtein}, {Bertin}, {Borsenberger}, {Copet}, {de Batz}, {Fouque},
  {Kimeswenger}, {Le Bertre}, {Lacombe}, {Rouan}, \& {Tiphene}}]{delfosse97}
{Delfosse}, X., {Tinney}, C.~G., {Forveille}, T., {et~al.} 1997, A\&A, 327, L25

\bibitem[{{Dhillon} {et~al.}(2021){Dhillon}, {Bezawada}, {Black}, {Dixon},
  {Gamble}, {Gao}, {Henry}, {Kerry}, {Littlefair}, {Lunney}, {Marsh}, {Miller},
  {Parsons}, {Ashley}, {Breedt}, {Brown}, {Dyer}, {Green}, {Pelisoli},
  {Sahman}, {Wild}, {Ives}, {Mehrgan}, {Stegmeier}, {Dubbeldam}, {Morris},
  {Osborn}, {Wilson}, {Casares}, {Mu{\~n}oz-Darias}, {Pall{\'e}},
  {Rodr{\'\i}guez-Gil}, {Shahbaz}, {Torres}, {de Ugarte Postigo},
  {Cabrera-Lavers}, {Corradi}, {Dom{\'\i}nguez}, \&
  {Garc{\'\i}a-Alvarez}}]{dhillon21a}
{Dhillon}, V.~S., {Bezawada}, N., {Black}, M., {et~al.} 2021, MNRAS, 507, 350

\bibitem[{{Digby} {et~al.}(2003){Digby}, {Hambly}, {Cooke}, {Reid}, \&
  {Cannon}}]{digby03}
{Digby}, A.~P., {Hambly}, N.~C., {Cooke}, J.~A., {Reid}, I.~N., \& {Cannon},
  R.~D. 2003, MNRAS, 344, 583

\bibitem[{{Dupuy} \& {Kraus}(2013)}]{dupuy13b}
{Dupuy}, T.~J. \& {Kraus}, A.~L. 2013, Science, 341, 1492

\bibitem[{{Dupuy} \& {Liu}(2012)}]{dupuy12}
{Dupuy}, T.~J. \& {Liu}, M.~C. 2012, ApJS, 201, 19

\bibitem[{{Eisenhardt} {et~al.}(2020){Eisenhardt}, {Marocco}, {Fowler},
  {Meisner}, {Kirkpatrick}, {Garcia}, {Jarrett}, {Koontz}, {Marchese},
  {Stanford}, {Caselden}, {Cushing}, {Cutri}, {Faherty}, {Gelino}, {Gonzalez},
  {Mainzer}, {Mobasher}, {Schlegel}, {Stern}, {Teplitz}, \&
  {Wright}}]{eisenhardt20}
{Eisenhardt}, P. R.~M., {Marocco}, F., {Fowler}, J.~W., {et~al.} 2020, \apjs,
  247, 69

\bibitem[{{Faherty} {et~al.}(2014){Faherty}, {Beletsky}, {Burgasser}, {Tinney},
  {Osip}, {Filippazzo}, \& {Simcoe}}]{faherty14a}
{Faherty}, J.~K., {Beletsky}, Y., {Burgasser}, A.~J., {et~al.} 2014, ApJ, 790,
  90

\bibitem[{{Faherty} {et~al.}(2012){Faherty}, {Burgasser}, {Walter}, {Van der
  Bliek}, {Shara}, {Cruz}, {West}, {Vrba}, \& {Anglada-Escud{\'e}}}]{faherty12}
{Faherty}, J.~K., {Burgasser}, A.~J., {Walter}, F.~M., {et~al.} 2012, ApJ, 752,
  56

\bibitem[{{Filippenko}(1982)}]{filippenko82}
{Filippenko}, A.~V. 1982, \pasp, 94, 715

\bibitem[{{Fritz} {et~al.}(2010){Fritz}, {Gillessen}, {Trippe}, {Ott},
  {Bartko}, {Pfuhl}, {Dodds-Eden}, {Davies}, {Eisenhauer}, \&
  {Genzel}}]{fritz10}
{Fritz}, T., {Gillessen}, S., {Trippe}, S., {et~al.} 2010, \mnras, 401, 1177

\bibitem[{{Fuchs} {et~al.}(2009){Fuchs}, {Dettbarn}, {Rix}, {Beers}, {Bizyaev},
  {Brewington}, {Jahrei{\ss}}, {Klement}, {Malanushenko}, {Malanushenko},
  {Oravetz}, {Pan}, {Simmons}, \& {Snedden}}]{fuchs09a}
{Fuchs}, B., {Dettbarn}, C., {Rix}, H.-W., {et~al.} 2009, AJ, 137, 4149

\bibitem[{{Garz{\'o}n} {et~al.}(2007){Garz{\'o}n}, {Abreu}, {Barrera},
  {Becerril}, {Cair{\'o}s}, {D{\'{\i}}az}, {Fragoso}, {Gago}, {Grange},
  {Gonz{\'a}lez}, {L{\'o}pez}, {Patr{\'o}n}, {P{\'e}rez}, {Rasilla}, {Redondo},
  {Restrepo}, {Saavedra}, {S{\'a}nchez}, {Tenegi}, \& {Vallb{\'e}}}]{garzon07}
{Garz{\'o}n}, F., {Abreu}, D., {Barrera}, S., {et~al.} 2007, in Revista
  Mexicana de Astronomia y Astrofisica Conference Series, Vol.~29, Revista
  Mexicana de Astronomia y Astrofisica Conference Series, ed. R.~{Guzm{\'a}n},
  12--17

\bibitem[{{Gerasimov} {et~al.}(2022){Gerasimov}, {Burgasser}, {Homeier},
  {Bedin}, {Rees}, {Scalco}, {Anderson}, \& {Salaris}}]{gerasimov22}
{Gerasimov}, R., {Burgasser}, A.~J., {Homeier}, D., {et~al.} 2022, ApJ, 930, 24

\bibitem[{{Gizis}(1997)}]{gizis97a}
{Gizis}, J.~E. 1997, AJ, 113, 806

\bibitem[{{Gonzales} {et~al.}(2021){Gonzales}, {Burningham}, {Faherty},
  {Visscher}, {Marley}, {Lupu}, {Freedman}, \& {Lewis}}]{gonzales21}
{Gonzales}, E.~C., {Burningham}, B., {Faherty}, J.~K., {et~al.} 2021, ApJ, 923,
  19

\bibitem[{{Goodman}(2021)}]{goodman21a}
{Goodman}, S.~J. 2021, Research Notes of the American Astronomical Society, 5,
  178

\bibitem[{{Greco} {et~al.}(2019){Greco}, {Schneider}, {Cushing}, {Kirkpatrick},
  \& {Burgasser}}]{greco19a}
{Greco}, J.~J., {Schneider}, A.~C., {Cushing}, M.~C., {Kirkpatrick}, J.~D., \&
  {Burgasser}, A.~J. 2019, AJ, 158, 182

\bibitem[{{Green}(1985)}]{green85}
{Green}, R.~M. 1985, {Spherical Astronomy}

\bibitem[{{Hambly} {et~al.}(2008){Hambly}, {Collins}, {Cross}, {Mann}, {Read},
  {Sutorius}, {Bond}, {Bryant}, {Emerson}, {Lawrence}, {Rimoldini}, {Stewart},
  {Williams}, {Adamson}, {Hirst}, {Dye}, \& {Warren}}]{hambly08}
{Hambly}, N.~C., {Collins}, R.~S., {Cross}, N.~J.~G., {et~al.} 2008, MNRAS,
  384, 637

\bibitem[{{Henry} {et~al.}(2006){Henry}, {Jao}, {Subasavage}, {Beaulieu},
  {Ianna}, {Costa}, \& {M{\'e}ndez}}]{henry06}
{Henry}, T.~J., {Jao}, W.-C., {Subasavage}, J.~P., {et~al.} 2006, AJ, 132, 2360

\bibitem[{{Henry} {et~al.}(2018){Henry}, {Jao}, {Winters}, {Dieterich},
  {Finch}, {Ianna}, {Riedel}, {Silverstein}, {Subasavage}, \&
  {Vrijmoet}}]{henry18}
{Henry}, T.~J., {Jao}, W.-C., {Winters}, J.~G., {et~al.} 2018, AJ, 155, 265

\bibitem[{{Hewett} {et~al.}(2006){Hewett}, {Warren}, {Leggett}, \&
  {Hodgkin}}]{hewett06}
{Hewett}, P.~C., {Warren}, S.~J., {Leggett}, S.~K., \& {Hodgkin}, S.~T. 2006,
  MNRAS, 367, 454

\bibitem[{{Hodgkin} {et~al.}(2009){Hodgkin}, {Irwin}, {Hewett}, \&
  {Warren}}]{hodgkin09}
{Hodgkin}, S.~T., {Irwin}, M.~J., {Hewett}, P.~C., \& {Warren}, S.~J. 2009,
  MNRAS, 394, 675

\bibitem[{{Houk} \& {Swift}(1999)}]{houk99}
{Houk}, N. \& {Swift}, C. 1999, Michigan Spectral Survey, 5, 0

\bibitem[{{Hsu} {et~al.}(2021){Hsu}, {Burgasser}, {Theissen}, {Gelino},
  {Birky}, {Diamant}, {Bardalez Gagliuffi}, {Aganze}, {Blake}, \&
  {Faherty}}]{hsu21}
{Hsu}, C.-C., {Burgasser}, A.~J., {Theissen}, C.~A., {et~al.} 2021, \apjs, 257,
  45

\bibitem[{{Irwin} {et~al.}(2004){Irwin}, {Lewis}, {Hodgkin}, {Bunclark},
  {Evans}, {McMahon}, {Emerson}, {Stewart}, \& {Beard}}]{irwin04}
{Irwin}, M.~J., {Lewis}, J., {Hodgkin}, S., {et~al.} 2004, in Optimizing
  Scientific Return for Astronomy through Information Technologies. Edited by
  Quinn, Peter J.; Bridger, Alan. Proceedings of the SPIE, Volume 5493, pp.
  411-422 (2004)., ed. P.~J. {Quinn} \& A.~{Bridger}, 411--422

\bibitem[{{Johnson} \& {Soderblom}(1987)}]{johnson87}
{Johnson}, D.~R.~H. \& {Soderblom}, D.~R. 1987, AJ, 93, 864

\bibitem[{{Juri{\'c}} {et~al.}(2008){Juri{\'c}}, {Ivezi{\'c}}, {Brooks},
  {Lupton}, {Schlegel}, {Finkbeiner}, {Padmanabhan}, {Bond}, {Sesar},
  {Rockosi}, {Knapp}, {Gunn}, {Sumi}, {Schneider}, {Barentine}, {Brewington},
  {Brinkmann}, {Fukugita}, {Harvanek}, {Kleinman}, {Krzesinski}, {Long},
  {Neilsen}, {Nitta}, {Snedden}, \& {York}}]{juric08a}
{Juri{\'c}}, M., {Ivezi{\'c}}, {\v{Z}}., {Brooks}, A., {et~al.} 2008, ApJ, 673,
  864

\bibitem[{{Kirkpatrick}(2005)}]{kirkpatrick05}
{Kirkpatrick}, J.~D. 2005, ARA\&A, 43, 195

\bibitem[{{Kirkpatrick} {et~al.}(1997){Kirkpatrick}, {Beichman}, \&
  {Skrutskie}}]{kirkpatrick97}
{Kirkpatrick}, J.~D., {Beichman}, C.~A., \& {Skrutskie}, M.~F. 1997, ApJ, 476,
  311

\bibitem[{{Kirkpatrick} {et~al.}(2011){Kirkpatrick}, {Cushing}, {Gelino},
  {Griffith}, {Skrutskie}, {Marsh}, {Wright}, {Mainzer}, {Eisenhardt},
  {McLean}, \& {30 co-authors}}]{kirkpatrick11}
{Kirkpatrick}, J.~D., {Cushing}, M.~C., {Gelino}, C.~R., {et~al.} 2011, ApJS,
  197, 19

\bibitem[{{Kirkpatrick} {et~al.}(2012){Kirkpatrick}, {Gelino}, {Cushing},
  {Mace}, {Griffith}, {Skrutskie}, {Marsh}, {Wright}, {Eisenhardt}, {McLean},
  {Mainzer}, {Burgasser}, {Tinney}, {Parker}, \& {Salter}}]{kirkpatrick12}
{Kirkpatrick}, J.~D., {Gelino}, C.~R., {Cushing}, M.~C., {et~al.} 2012, ApJ,
  753, 156

\bibitem[{{Kirkpatrick} {et~al.}(2021{\natexlab{a}}){Kirkpatrick}, {Gelino},
  {Faherty}, {Meisner}, {Caselden}, {Schneider}, {Marocco}, {Cayago}, {Smart},
  {Eisenhardt}, {Kuchner}, {Wright}, {Cushing}, {Allers}, {Bardalez Gagliuffi},
  {Burgasser}, {Gagn{\'e}}, {Logsdon}, {Martin}, {Ingalls}, {Lowrance},
  {Abrahams}, {Aganze}, {Gerasimov}, {Gonzales}, {Hsu}, {Kamraj}, {Kiman},
  {Rees}, {Theissen}, {Ammar}, {Andersen}, {Beaulieu}, {Colin}, {Elachi},
  {Goodman}, {Gramaize}, {Hamlet}, {Hong}, {Jonkeren}, {Khalil}, {Martin},
  {Pendrill}, {Pumphrey}, {Rothermich}, {Sainio}, {Stenner}, {Tanner},
  {Th{\'e}venot}, {Voloshin}, {Walla}, {W{\k{e}}dracki}, \& {Backyard Worlds:
  Planet 9 Collaboration}}]{kirkpatrick21c}
{Kirkpatrick}, J.~D., {Gelino}, C.~R., {Faherty}, J.~K., {et~al.}
  2021{\natexlab{a}}, \apjs, 253, 7

\bibitem[{{Kirkpatrick} {et~al.}(2021{\natexlab{b}}){Kirkpatrick}, {Marocco},
  {Caselden}, {Meisner}, {Faherty}, {Schneider}, {Kuchner}, {Casewell},
  {Gelino}, {Cushing}, {Eisenhardt}, {Wright}, \& {Schurr}}]{kirkpatrick21b}
{Kirkpatrick}, J.~D., {Marocco}, F., {Caselden}, D., {et~al.}
  2021{\natexlab{b}}, ApJL, 915, L6

\bibitem[{{Kirkpatrick} {et~al.}(2019){Kirkpatrick}, {Martin}, {Smart},
  {Cayago}, {Beichman}, {Marocco}, {Gelino}, {Faherty}, {Cushing}, {Schneider},
  {Mace}, {Tinney}, {Wright}, {Lowrance}, {Ingalls}, {Vrba}, {Munn}, {Dahm}, \&
  {McLean}}]{kirkpatrick19}
{Kirkpatrick}, J.~D., {Martin}, E.~C., {Smart}, R.~L., {et~al.} 2019, ApJS,
  240, 19

\bibitem[{{Kirkpatrick} {et~al.}(2014){Kirkpatrick}, {Schneider},
  {Fajardo-Acosta}, {Gelino}, {Mace}, {Wright}, {Logsdon}, {McLean}, {Cushing},
  {Skrutskie}, {Eisenhardt}, {Stern}, {Balokovi{\'c}}, {Burgasser}, {Faherty},
  {Lansbury}, {Rich}, {Skrzypek}, {Fowler}, {Cutri}, {Masci}, {Conrow},
  {Grillmair}, {McCallon}, {Beichman}, \& {Marsh}}]{kirkpatrick14}
{Kirkpatrick}, J.~D., {Schneider}, A., {Fajardo-Acosta}, S., {et~al.} 2014,
  ApJ, 783, 122

\bibitem[{{Kov\'acs} {et~al.}(2004){Kov\'acs}, {Mall}, {Bizenberger},
  {Baumeister}, \& {R{\"o}ser}}]{kovacs04}
{Kov\'acs}, Z., {Mall}, U., {Bizenberger}, P., {Baumeister}, H., \&
  {R{\"o}ser}, H. 2004, in Society of Photo-Optical Instrumentation Engineers
  (SPIE) Conference Series, Vol. 5499, Society of Photo-Optical Instrumentation
  Engineers (SPIE) Conference Series, ed. {J.~D.~Garnett \& J.~W.~Beletic},
  432--441

\bibitem[{{L{\' e}pine} {et~al.}(2003){L{\' e}pine}, {Rich}, \&
  {Shara}}]{lepine03a}
{L{\' e}pine}, S., {Rich}, R.~M., \& {Shara}, M.~M. 2003, AJ, 125, 1598

\bibitem[{{Lawrence} {et~al.}(2007){Lawrence}, {Warren}, {Almaini}, {Edge},
  {Hambly}, \& {17 co-authors}}]{lawrence07}
{Lawrence}, A., {Warren}, S.~J., {Almaini}, O., {et~al.} 2007, MNRAS, 379, 1599

\bibitem[{{Leggett} {et~al.}(2012){Leggett}, {Saumon}, {Marley}, {Lodders},
  {Canty}, {Lucas}, {Smart}, {Tinney}, {Homeier}, {Allard}, {Burningham},
  {Day-Jones}, {Fegley}, {Ishii}, {Jones}, {Marocco}, {Pinfield}, \&
  {Tamura}}]{leggett12b}
{Leggett}, S.~K., {Saumon}, D., {Marley}, M.~S., {et~al.} 2012, \apj, 748, 74

\bibitem[{{Leggett} {et~al.}(2017){Leggett}, {Tremblin}, {Esplin}, {Luhman}, \&
  {Morley}}]{leggett17}
{Leggett}, S.~K., {Tremblin}, P., {Esplin}, T.~L., {Luhman}, K.~L., \&
  {Morley}, C.~V. 2017, \apj, 842, 118

\bibitem[{{L{\'e}pine} \& {Gaidos}(2013)}]{lepine13a}
{L{\'e}pine}, S. \& {Gaidos}, E. 2013, Astron. Nachr., 334, 176

\bibitem[{{L{\'e}pine} {et~al.}(2007){L{\'e}pine}, {Rich}, \&
  {Shara}}]{lepine07c}
{L{\'e}pine}, S., {Rich}, R.~M., \& {Shara}, M.~M. 2007, ApJ, 669, 1235

\bibitem[{{Lodieu}(2018)}]{lodieu18c}
{Lodieu}, N. 2018, {Metal-Depleted Brown Dwarfs}, ed. H.~J. {Deeg} \& J.~A.
  {Belmonte}, 173

\bibitem[{{Lodieu} {et~al.}(2019{\natexlab{a}}){Lodieu}, {Allard}, {Rodrigo},
  {Pavlenko}, {Burgasser}, {Lyubchik}, {Kaminsky}, \& {Homeier}}]{lodieu19c}
{Lodieu}, N., {Allard}, F., {Rodrigo}, C., {et~al.} 2019{\natexlab{a}}, A\&A,
  628, A61

\bibitem[{{Lodieu} {et~al.}(2012){Lodieu}, {Espinoza Contreras}, {Zapatero
  Osorio}, {Solano}, {Aberasturi}, \& {Mart{\'{\i}}n}}]{lodieu12b}
{Lodieu}, N., {Espinoza Contreras}, M., {Zapatero Osorio}, M.~R., {et~al.}
  2012, A\&A, 542, A105

\bibitem[{{Lodieu} {et~al.}(2017){Lodieu}, {Espinoza Contreras}, {Zapatero
  Osorio}, {Solano}, {Aberasturi}, {Mart{\'{\i}}n}, \& {Rodrigo}}]{lodieu17a}
{Lodieu}, N., {Espinoza Contreras}, M., {Zapatero Osorio}, M.~R., {et~al.}
  2017, A\&A, 598, A92

\bibitem[{{Lodieu} {et~al.}(2019{\natexlab{b}}){Lodieu}, {Smart},
  {P{\'e}rez-Garrido}, \& {Silvotti}}]{lodieu19a}
{Lodieu}, N., {Smart}, R.~L., {P{\'e}rez-Garrido}, A., \& {Silvotti}, R.
  2019{\natexlab{b}}, A\&A, 623, A35

\bibitem[{{Lodieu} {et~al.}(2015){Lodieu}, {Zapatero Osorio}, {Rebolo},
  {B{\'e}jar}, {Pavlenko}, \& {P{\'e}rez-Garrido}}]{lodieu15b}
{Lodieu}, N., {Zapatero Osorio}, M.~R., {Rebolo}, R., {et~al.} 2015, A\&A, 581,
  A73

\bibitem[{{Looper} {et~al.}(2007){Looper}, {Kirkpatrick}, \&
  {Burgasser}}]{looper07}
{Looper}, D.~L., {Kirkpatrick}, J.~D., \& {Burgasser}, A.~J. 2007, AJ, 134,
  1162

\bibitem[{{Lucas} {et~al.}(2008){Lucas}, {Hoare}, {Longmore}, {Schr{\"o}der},
  \& {27 co-authors}}]{lucas08}
{Lucas}, P.~W., {Hoare}, M.~G., {Longmore}, A., {Schr{\"o}der}, A.~C., \& {27
  co-authors}. 2008, MNRAS, 391, 136

\bibitem[{{Luhman}(2013)}]{luhman13a}
{Luhman}, K.~L. 2013, ApJL, 767, L1

\bibitem[{{Luhman}(2014)}]{luhman14b}
{Luhman}, K.~L. 2014, ApJL, 786, L18

\bibitem[{{Luhman} \& {Esplin}(2014)}]{luhman14d}
{Luhman}, K.~L. \& {Esplin}, T.~L. 2014, ApJ, 796, 6

\bibitem[{{Mace} {et~al.}(2013){Mace}, {Kirkpatrick}, {Cushing}, {Gelino},
  {Griffith}, {Skrutskie}, {Marsh}, {Wright}, {Eisenhardt}, {McLean},
  {Thompson}, {Mix}, {Bailey}, {Beichman}, {Bloom}, {Burgasser}, {Fortney},
  {Hinz}, {Knox}, {Lowrance}, {Marley}, {Morley}, {Rodigas}, {Saumon},
  {Sheppard}, \& {Stock}}]{mace13}
{Mace}, G.~N., {Kirkpatrick}, J.~D., {Cushing}, M.~C., {et~al.} 2013, ApJS,
  205, 6

\bibitem[{{Mainzer} {et~al.}(2011){Mainzer}, {Cushing}, {Skrutskie}, {Gelino},
  {Kirkpatrick}, {Jarrett}, {Masci}, \& {13 co-authors}}]{mainzer11}
{Mainzer}, A., {Cushing}, M.~C., {Skrutskie}, M., {et~al.} 2011, ApJ, 726, 30

\bibitem[{{Marfil} {et~al.}(2021){Marfil}, {Tabernero}, {Montes}, {Caballero},
  {L{\'a}zaro}, {Gonz{\'a}lez Hern{\'a}ndez}, {Nagel}, {Passegger},
  {Schweitzer}, {Ribas}, {Reiners}, {Quirrenbach}, {Amado}, {Cifuentes},
  {Cort{\'e}s-Contreras}, {Dreizler}, {Duque-Arribas},
  {Galad{\'\i}-Enr{\'\i}quez}, {Henning}, {Jeffers}, {Kaminski}, {K{\"u}rster},
  {Lafarga}, {L{\'o}pez-Gallifa}, {Morales}, {Shan}, \&
  {Zechmeister}}]{marfil21}
{Marfil}, E., {Tabernero}, H.~M., {Montes}, D., {et~al.} 2021, A\&A, 656, A162

\bibitem[{{Mart\'{\i}n}(1997)}]{martin97a}
{Mart\'{\i}n}, E.~L. 1997, A\&A, 321, 492

\bibitem[{{Mart{\'{\i}}n} {et~al.}(2018){Mart{\'{\i}}n}, {Lodieu}, {Pavlenko},
  \& {B{\'e}jar}}]{martin18a}
{Mart{\'{\i}}n}, E.~L., {Lodieu}, N., {Pavlenko}, Y., \& {B{\'e}jar}, V.~J.~S.
  2018, ApJ, 856, 40

\bibitem[{{Meisner} {et~al.}(2020{\natexlab{a}}){Meisner}, {Caselden},
  {Kirkpatrick}, {Marocco}, {Gelino}, {Cushing}, {Eisenhardt}, {Wright},
  {Faherty}, {Koontz}, {Marchese}, {Khalil}, {Fowler}, \&
  {Schlafly}}]{meisner20b}
{Meisner}, A.~M., {Caselden}, D., {Kirkpatrick}, J.~D., {et~al.}
  2020{\natexlab{a}}, ApJ, 889, 74

\bibitem[{{Meisner} {et~al.}(2020{\natexlab{b}}){Meisner}, {Faherty},
  {Kirkpatrick}, {Schneider}, {Caselden}, {Gagn{\'e}}, {Kuchner}, {Burgasser},
  {Casewell}, {Debes}, {Artigau}, {Bardalez Gagliuffi}, {Logsdon}, {Kiman},
  {Allers}, {Hsu}, {Wisniewski}, {Allen}, {Beaulieu}, {Colin}, {Durantini
  Luca}, {Goodman}, {Gramaize}, {Hamlet}, {Hinckley}, {Kiwy}, {Martin},
  {Pendrill}, {Rothermich}, {Sainio}, {Sch{\"u}mann}, {Andersen}, {Tanner},
  {Thakur}, {Th{\'e}venot}, {Walla}, {W{\k{e}}dracki}, {Aganze}, {Gerasimov},
  {Theissen}, \& {Backyard Worlds: Planet 9 Collaboration}}]{meisner20a}
{Meisner}, A.~M., {Faherty}, J.~K., {Kirkpatrick}, J.~D., {et~al.}
  2020{\natexlab{b}}, ApJ, 899, 123

\bibitem[{{Meisner} {et~al.}(2021){Meisner}, {Schneider}, {Burgasser},
  {Marocco}, {Line}, {Faherty}, {Kirkpatrick}, {Caselden}, {Kuchner}, {Gelino},
  {Gagn{\'e}}, {Theissen}, {Gerasimov}, {Aganze}, {Hsu}, {Wisniewski},
  {Casewell}, {Bardalez Gagliuffi}, {Logsdon}, {Eisenhardt}, {Allers}, {Debes},
  {Allen}, {Stevnbak Andersen}, {Goodman}, {Gramaize}, {Martin}, {Sainio},
  {Cushing}, \& {Backyard Worlds: Planet 9 Collaboration}}]{meisner21}
{Meisner}, A.~M., {Schneider}, A.~C., {Burgasser}, A.~J., {et~al.} 2021, \apj,
  915, 120

\bibitem[{{Minniti} {et~al.}(2010){Minniti}, {Lucas}, {Emerson}, {Saito},
  {Hempel}, {Pietrukowicz}, {Ahumada}, {Alonso}, {Alonso-Garcia}, {Arias},
  {Bandyopadhyay}, {Barb{\'a}}, {Barbuy}, {Bedin}, {Bica}, {Borissova},
  {Bronfman}, {Carraro}, {Catelan}, {Clari{\'a}}, {Cross}, {de Grijs},
  {D{\'e}k{\'a}ny}, {Drew}, {Fari{\~n}a}, {Feinstein}, {Fern{\'a}ndez
  Laj{\'u}s}, {Gamen}, {Geisler}, {Gieren}, {Goldman}, {Gonzalez}, {Gunthardt},
  {Gurovich}, {Hambly}, {Irwin}, {Ivanov}, {Jord{\'a}n}, {Kerins}, {Kinemuchi},
  {Kurtev}, {L{\'o}pez-Corredoira}, {Maccarone}, {Masetti}, {Merlo},
  {Messineo}, {Mirabel}, {Monaco}, {Morelli}, {Padilla}, {Palma}, {Parisi},
  {Pignata}, {Rejkuba}, {Roman-Lopes}, {Sale}, {Schreiber}, {Schr{\"o}der},
  {Smith}, {}, {Soto}, {Tamura}, {Tappert}, {Thompson}, {Toledo}, {Zoccali}, \&
  {Pietrzynski}}]{minniti10}
{Minniti}, D., {Lucas}, P.~W., {Emerson}, J.~P., {et~al.} 2010, Nature
  Astronomy, 15, 433

\bibitem[{{Monet} {et~al.}(1992){Monet}, {Dahn}, {Vrba}, {Harris}, {Pier},
  {Luginbuhl}, \& {Ables}}]{monet92}
{Monet}, D.~G., {Dahn}, C.~C., {Vrba}, F.~J., {et~al.} 1992, AJ, 103, 638

\bibitem[{{Murray} {et~al.}(2011){Murray}, {Burningham}, {Jones}, {Pinfield},
  {Lucas}, {Leggett}, {Tinney}, {Day-Jones}, {Weights}, {Lodieu}, {P{\'e}rez
  Prieto}, {Nickson}, {Zhang}, {Clarke}, {Jenkins}, \& {Tamura}}]{murray11}
{Murray}, D.~N., {Burningham}, B., {Jones}, H.~R.~A., {et~al.} 2011, MNRAS,
  414, 575

\bibitem[{{Nakajima} {et~al.}(1995){Nakajima}, {Oppenheimer}, {Kulkarni},
  {Golimowski}, {Matthews}, \& {Durrance}}]{nakajima95}
{Nakajima}, T., {Oppenheimer}, B.~R., {Kulkarni}, S.~R., {et~al.} 1995, Nature,
  378, 463

\bibitem[{{Passegger} {et~al.}(2018){Passegger}, {Reiners}, {Jeffers},
  {Wende-von Berg}, {Sch{\"o}fer}, {Caballero}, {Schweitzer}, {Amado},
  {B{\'e}jar}, {Cort{\'e}s-Contreras}, {Hatzes}, {K{\"u}rster}, {Montes},
  {Pedraz}, {Quirrenbach}, {Ribas}, \& {Seifert}}]{passegger18}
{Passegger}, V.~M., {Reiners}, A., {Jeffers}, S.~V., {et~al.} 2018, A\&A, 615,
  A6

\bibitem[{{Pinfield} {et~al.}(2012){Pinfield}, {Burningham}, {Lodieu},
  {Leggett}, {Tinney}, {van Spaandonk}, {Marocco}, {Smart}, {Gomes}, {Smith},
  {Lucas}, {Day-Jones}, {Murray}, {Katsiyannis}, {Catalan}, {Cardoso},
  {Clarke}, {Folkes}, {G{\'a}lvez-Ortiz}, {Homeier}, {Jenkins}, {Jones}, \&
  {Zhang}}]{pinfield12}
{Pinfield}, D.~J., {Burningham}, B., {Lodieu}, N., {et~al.} 2012, MNRAS, 422,
  1922

\bibitem[{{Quirrenbach} {et~al.}(2020){Quirrenbach}, {CARMENES Consortium},
  {Amado}, {Ribas}, {Reiners}, {Caballero}, {Aceituno}, {Alacid},
  {Alonso-Floriano}, {Anglada-Escud{\'e}}, {Azzaro}, {Baroch}, {Bauer},
  {Becerril}, {B{\'e}jar}, {Bluhm}, {Calvo Ortega}, {Cardona Guill{\'e}n},
  {Casasayas-Barris}, {Chaturvedi}, {Cifuentes}, {Colom{\'e}}, {Conte},
  {Cort{\'e}s-Contreras}, {Czesla}, {D{\'\i}ez-Alonso}, {Dom{\'\i}nguez
  Fern{\'a}ndez}, {Dreizler}, {Duque-Arribas}, {Espinoza}, {Fuhrmeister},
  {Galad{\'\i}-Enr{\'\i}quez}, {Gar{\textasciiacute}a Quintana},
  {Gonz{\'a}lez-Alvare}, {Gonz{\'a}lez Cuesta}, {Gonz{\'a}lez Hern{\'a}ndez},
  {Guenther}, {de Guindos}, {Hatzes}, {Henning}, {Herbort}, {Herrero}, {Hintz},
  {Iglesias-P{\'a}ra}, {Jeffers}, {Johnson}, {de Juan}, {Kaminski}, {Kemmer},
  {Khaimova}, {Khalafinejad}, {Klahr}, {Kossakowski}, {Kreidberg},
  {K{\"u}rster}, {Labarga}, {Lafarga}, {Lamp{\'o}n}, {Lara}, {Lillo-Box},
  {Lodieu}, {L{\'o}pez Gallifa}, {L{\'o}pez Gonz{\'a}lez}, {L{\'o}pez-Puertas},
  {Luque}, {Marfil}, {Mart{\'\i}n-Ruiz}, {Matth{\'e}}, {Molaverdikhani},
  {Montes}, {Morales}, {Morales-Calder{\'o}on}, {Nagel}, {Nortmann}, {Nowak},
  {Ofir}, {Oshaghi}, {Pall{\'e}}, {Passegger}, {Pavlov}, {Pedraz},
  {Perdelwitz}, {Perger}, {Reffert}, {Revilla}, {Rodr{\'\i}guez},
  {Rodr{\'\i}guez L{\'o}pez}, {Sabotta}, {Sadegi}, {Sairam}, {Salz},
  {S{\'a}nchez-L{\'o}pez}, {Sanz-Forcada}, {Sarkis}, {Sch{\"a}fer}, {Schiller},
  {Schlecker}, {Schmitt}, {Sch{\"o}fer}, {Schweitzer}, {Seiferta}, {Shan},
  {Shulyak}, {Skrzypinski}, {Solano}, {Soto}, {Stahl}, {Stangret}, {Stock},
  {Strachan}, {Stuber}, {St{\"u}rmer}, {Tabernero}, {Tal-Or}, {Tala-Pinto},
  {Trifonov}, {Vanaverbeke}, {Yan}, {Zapatero Osorio}, \&
  {Zechmeister}}]{quirrenbach20}
{Quirrenbach}, A., {CARMENES Consortium}, {Amado}, P.~J., {et~al.} 2020, in
  Society of Photo-Optical Instrumentation Engineers (SPIE) Conference Series,
  Vol. 11447, Society of Photo-Optical Instrumentation Engineers (SPIE)
  Conference Series, 114473C

\bibitem[{{Rajpurohit} {et~al.}(2016){Rajpurohit}, {Reyl{\'e}}, {Allard},
  {Homeier}, {Bayo}, {Mousis}, {Rajpurohit}, \&
  {Fern{\'a}ndez-Trincado}}]{rajpurohit16a}
{Rajpurohit}, A.~S., {Reyl{\'e}}, C., {Allard}, F., {et~al.} 2016, A\&A, 596,
  A33

\bibitem[{{Rajpurohit} {et~al.}(2014){Rajpurohit}, {Reyl{\'e}}, {Allard},
  {Scholz}, {Homeier}, {Schultheis}, \& {Bayo}}]{rajpurohit14}
{Rajpurohit}, A.~S., {Reyl{\'e}}, C., {Allard}, F., {et~al.} 2014, A\&A, 564,
  A90

\bibitem[{{Rebolo} {et~al.}(1995){Rebolo}, {Zapatero-Osorio}, \&
  {Mart\'{\i}n}}]{rebolo95}
{Rebolo}, R., {Zapatero-Osorio}, M.~R., \& {Mart\'{\i}n}, E.~L. 1995, Nature,
  377, 129

\bibitem[{{Reddy} {et~al.}(2006){Reddy}, {Lambert}, \& {Allende
  Prieto}}]{reddy06}
{Reddy}, B.~E., {Lambert}, D.~L., \& {Allende Prieto}, C. 2006, MNRAS, 367,
  1329

\bibitem[{{Reyl{\'e}}(2018)}]{reyle18}
{Reyl{\'e}}, C. 2018, A\&A, 619, L8

\bibitem[{{Reyl{\'e}} {et~al.}(2021){Reyl{\'e}}, {Jardine}, {Fouqu{\'e}},
  {Caballero}, {Smart}, \& {Sozzetti}}]{reyle21}
{Reyl{\'e}}, C., {Jardine}, K., {Fouqu{\'e}}, P., {et~al.} 2021, A\&A, 650,
  A201

\bibitem[{{Robin} {et~al.}(2003){Robin}, {Reyl{\'e}}, {Derri{\`e}re}, \&
  {Picaud}}]{robin03}
{Robin}, A.~C., {Reyl{\'e}}, C., {Derri{\`e}re}, S., \& {Picaud}, S. 2003,
  A\&A, 409, 523

\bibitem[{{Rodrigo} \& {Solano}(2020)}]{rodrigo20}
{Rodrigo}, C. \& {Solano}, E. 2020, in XIV.0 Scientific Meeting (virtual) of
  the Spanish Astronomical Society, 182

\bibitem[{{Sahlmann} {et~al.}(2016){Sahlmann}, {Lazorenko}, {Bouy},
  {Mart{\'\i}n}, {Queloz}, {S{\'e}gransan}, \& {Zapatero Osorio}}]{sahlmann16}
{Sahlmann}, J., {Lazorenko}, P.~F., {Bouy}, H., {et~al.} 2016, \mnras, 455, 357

\bibitem[{{Saito} {et~al.}(2010){Saito}, {Hempel}, {Alonso-Garc{\'\i}a},
  {Toledo}, {Borissova}, {Gonz{\'a}lez}, {Beamin}, {Minniti}, {Lucas},
  {Emerson}, {Ahumada}, {Aigrain}, {Alonso}, {Am{\^o}res}, {Angeloni}, {Arias},
  {Bandyopadhyay}, {Barb{\'a}}, {Barbuy}, {Baume}, {Bedin}, {Bica}, {Bronfman},
  {Carraro}, {Catelan}, {Clari{\'a}}, {Contreras}, {Cross}, {Davis}, {de
  Grijs}, {D{\'e}k{\'a}ny}, {Janet Drew}, {Fari{\~n}a}, {Feinstein},
  {Fern{\'a}ndez Laj{\'u}s}, {Folkes}, {Gamen}, {Geisler}, {Gieren}, {Goldman},
  {Gosling}, {Gunthardt}, {Gurovich}, {Hambly}, {Hanson}, {Hoare}, {Irwin},
  {Ivanov}, {Jord{\'a}n}, {Kerins}, {Kinemuchi}, {Kurtev}, {Longmore},
  {L{\'o}pez-Corredoira}, {Maccarone}, {Mart{\'\i}n}, {Masetti}, {Mennickent},
  {Merlo}, {Messineo}, {Mirabel}, {Monaco}, {Moni Bidin}, {Morelli}, {Padilla},
  {Palma}, {Parisi}, {Parker}, {Pavani}, {Pietrukowicz}, {Pietrzynski},
  {Pignata}, {Rejkuba}, {Rojas}, {Roman Lopes}, {Ruiz}, {Sale}, {Saviane},
  {Schreiber}, {Schr{\"o}der}, {Sharma}, {Smith}, {Sodr{\'e}}, {Soto},
  {Stephens}, {Tamura}, {Tappert}, {Thompson}, {Valenti}, {Vanzi}, {Weidmann},
  \& {Zoccali}}]{saito10a}
{Saito}, R., {Hempel}, M., {Alonso-Garc{\'\i}a}, J., {et~al.} 2010, The
  Messenger, 141, 24

\bibitem[{{Saito} {et~al.}(2012){Saito}, {Hempel}, {Minniti}, {Lucas},
  {Rejkuba}, {Toledo}, {Gonzalez}, {Alonso-Garc{\'\i}a}, {Irwin},
  {Gonzalez-Solares}, {Hodgkin}, {Lewis}, {Cross}, {Ivanov}, {Kerins},
  {Emerson}, {Soto}, {Am{\^o}res}, {Gurovich}, {D{\'e}k{\'a}ny}, {Angeloni},
  {Beamin}, {Catelan}, {Padilla}, {Zoccali}, {Pietrukowicz}, {Moni Bidin},
  {Mauro}, {Geisler}, {Folkes}, {Sale}, {Borissova}, {Kurtev}, {Ahumada},
  {Alonso}, {Adamson}, {Arias}, {Bandyopadhyay}, {Barb{\'a}}, {Barbuy},
  {Baume}, {Bedin}, {Bellini}, {Benjamin}, {Bica}, {Bonatto}, {Bronfman},
  {Carraro}, {Chen{\`e}}, {Clari{\'a}}, {Clarke}, {Contreras}, {Corvill{\'o}n},
  {de Grijs}, {Dias}, {Drew}, {Fari{\~n}a}, {Feinstein},
  {Fern{\'a}ndez-Laj{\'u}s}, {Gamen}, {Gieren}, {Goldman},
  {Gonz{\'a}lez-Fern{\'a}ndez}, {Grand}, {Gunthardt}, {Hambly}, {Hanson},
  {He{\l}miniak}, {Hoare}, {Huckvale}, {Jord{\'a}n}, {Kinemuchi}, {Longmore},
  {L{\'o}pez-Corredoira}, {Maccarone}, {Majaess}, {Mart{\'\i}n}, {Masetti},
  {Mennickent}, {Mirabel}, {Monaco}, {Morelli}, {Motta}, {Palma}, {Parisi},
  {Parker}, {Pe{\~n}aloza}, {Pietrzy{\'n}ski}, {Pignata}, {Popescu}, {Read},
  {Rojas}, {Roman-Lopes}, {Ruiz}, {Saviane}, {Schreiber}, {Schr{\"o}der},
  {Sharma}, {Smith}, {Sodr{\'e}}, {Stead}, {Stephens}, {Tamura}, {Tappert},
  {Thompson}, {Valenti}, {Vanzi}, {Walton}, {Weidmann}, \&
  {Zijlstra}}]{saito12}
{Saito}, R.~K., {Hempel}, M., {Minniti}, D., {et~al.} 2012, \aap, 537, A107

\bibitem[{{Salvatier} {et~al.}(2016){Salvatier}, {Wiecki}, \&
  {Fonnesbeck}}]{salvatier16}
{Salvatier}, J., {Wiecki}, T.~V., \& {Fonnesbeck}, C. 2016, {PeerJ} Computer
  Science, 2, e55

\bibitem[{{Savcheva} {et~al.}(2014){Savcheva}, {West}, \&
  {Bochanski}}]{savcheva14}
{Savcheva}, A.~S., {West}, A.~A., \& {Bochanski}, J.~J. 2014, ApJ, 794, 145

\bibitem[{{Schmidt} {et~al.}(2010){Schmidt}, {West}, {Burgasser}, {Bochanski},
  \& {Hawley}}]{schmidt10a}
{Schmidt}, S.~J., {West}, A.~A., {Burgasser}, A.~J., {Bochanski}, J.~J., \&
  {Hawley}, S.~L. 2010, AJ, 139, 1045

\bibitem[{{Schneider} {et~al.}(2020){Schneider}, {Burgasser}, {Gerasimov},
  {Marocco}, {Gagn{\'e}}, {Goodman}, {Beaulieu}, {Pendrill}, {Rothermich},
  {Sainio}, {Kuchner}, {Caselden}, {Meisner}, {Faherty}, {Mamajek}, {Hsu},
  {Greco}, {Cushing}, {Kirkpatrick}, {Bardalez-Gagliuffi}, {Logsdon}, {Allers},
  {Debes}, \& {Backyard Worlds: Planet 9 Collaboration}}]{schneider20a}
{Schneider}, A.~C., {Burgasser}, A.~J., {Gerasimov}, R., {et~al.} 2020, ApJ,
  898, 77

\bibitem[{{Scholz}(2010)}]{scholz10a}
{Scholz}, R.-D. 2010, A\&A, 510, L8

\bibitem[{{Skrzypek} {et~al.}(2015){Skrzypek}, {Warren}, {Faherty}, {Mortlock},
  {Burgasser}, \& {Hewett}}]{skrzypek15}
{Skrzypek}, N., {Warren}, S.~J., {Faherty}, J.~K., {et~al.} 2015, A\&A, 574,
  A78

\bibitem[{{Stone}(2002)}]{stone02}
{Stone}, R.~C. 2002, PASP, 114, 1070

\bibitem[{{Tinney} {et~al.}(2005){Tinney}, {Burgasser}, {Kirkpatrick}, \&
  {McElwain}}]{tinney05}
{Tinney}, C.~G., {Burgasser}, A.~J., {Kirkpatrick}, J.~D., \& {McElwain}, M.~W.
  2005, AJ, 130, 2326

\bibitem[{{Tinney} {et~al.}(2014){Tinney}, {Faherty}, {Kirkpatrick}, {Cushing},
  {Morley}, \& {Wright}}]{tinney14}
{Tinney}, C.~G., {Faherty}, J.~K., {Kirkpatrick}, J.~D., {et~al.} 2014, ApJ,
  796, 39

\bibitem[{{Tody}(1986)}]{tody86}
{Tody}, D. 1986, in Society of Photo-Optical Instrumentation Engineers (SPIE)
  Conference Series, Vol. 627, Society of Photo-Optical Instrumentation
  Engineers (SPIE) Conference Series, ed. D.~L. {Crawford}, 733

\bibitem[{{Tody}(1993)}]{tody93}
{Tody}, D. 1993, in Astronomical Society of the Pacific Conference Series,
  Vol.~52, Astronomical Data Analysis Software and Systems II, ed. R.~J.
  {Hanisch}, R.~J.~V. {Brissenden}, \& J.~{Barnes}, 173

\bibitem[{{Vrba} {et~al.}(2004){Vrba}, {Henden}, {Luginbuhl}, {Guetter},
  {Munn}, {Canzian}, {Burgasser}, {Kirkpatrick}, {Fan}, {Geballe},
  {Golimowski}, {Knapp}, {Leggett}, {Schneider}, \& {Brinkmann}}]{vrba04}
{Vrba}, F.~J., {Henden}, A.~A., {Luginbuhl}, C.~B., {et~al.} 2004, AJ, 127,
  2948

\bibitem[{{Wesemael} {et~al.}(1993){Wesemael}, {Greenstein}, {Liebert},
  {Lamontagne}, {Fontaine}, {Bergeron}, \& {Glaspey}}]{Wesemael93}
{Wesemael}, F., {Greenstein}, J.~L., {Liebert}, J., {et~al.} 1993, PASP, 105,
  761

\bibitem[{{Wright} {et~al.}(2010){Wright}, {Eisenhardt}, {Mainzer}, {Ressler},
  {Cutri}, {Jarrett}, {Kirkpatrick}, \& {31 co-authors}}]{wright10}
{Wright}, E.~L., {Eisenhardt}, P.~R.~M., {Mainzer}, A.~K., {et~al.} 2010, AJ,
  140, 1868

\bibitem[{{Zapatero Osorio} {et~al.}(2016){Zapatero Osorio}, {Lodieu},
  {B{\'e}jar}, {Mart{\'\i}n}, {Ivanov}, {Bayo}, {Boffin}, {Mu{\v{z}}i{\'c}},
  {Minniti}, \& {Beam{\'\i}n}}]{zapatero16}
{Zapatero Osorio}, M.~R., {Lodieu}, N., {B{\'e}jar}, V.~J.~S., {et~al.} 2016,
  \aap, 592, A80

\bibitem[{{Zapatero Osorio} {et~al.}(2006){Zapatero Osorio}, {Mart{\'{\i}}n},
  {Bouy}, {Tata}, {Deshpande}, \& {Wainscoat}}]{zapatero06}
{Zapatero Osorio}, M.~R., {Mart{\'{\i}}n}, E.~L., {Bouy}, H., {et~al.} 2006,
  ApJ, 647, 1405

\bibitem[{{Zhang} {et~al.}(2021){Zhang}, {Liu}, {Marley}, {Line}, \&
  {Best}}]{zhang21}
{Zhang}, Z., {Liu}, M.~C., {Marley}, M.~S., {Line}, M.~R., \& {Best}, W. M.~J.
  2021, \apj, 921, 95

\bibitem[{{Zhang} {et~al.}(2018){Zhang}, {Galvez-Ortiz}, {Pinfield},
  {Burgasser}, {Lodieu}, {Jones}, {Mart{\'\i}n}, {Burningham}, {Homeier},
  {Allard}, {Zapatero Osorio}, {Smith}, {Smart}, {L{\'o}pez Mart{\'\i}},
  {Marocco}, \& {Rebolo}}]{zhang18b}
{Zhang}, Z.~H., {Galvez-Ortiz}, M.~C., {Pinfield}, D.~J., {et~al.} 2018, MNRAS,
  480, 5447

\bibitem[{{Zhang} {et~al.}(2017{\natexlab{a}}){Zhang}, {Homeier}, {Pinfield},
  {Lodieu}, {Jones}, {Allard}, \& {Pavlenko}}]{zhang17b}
{Zhang}, Z.~H., {Homeier}, D., {Pinfield}, D.~J., {et~al.} 2017{\natexlab{a}},
  MNRAS, 468, 261

\bibitem[{{Zhang} {et~al.}(2017{\natexlab{b}}){Zhang}, {Pinfield},
  {G{\'a}lvez-Ortiz}, {Burningham}, {Lodieu}, {Marocco}, {Burgasser},
  {Day-Jones}, {Allard}, {Jones}, {Homeier}, {Gomes}, \& {Smart}}]{zhang17a}
{Zhang}, Z.~H., {Pinfield}, D.~J., {G{\'a}lvez-Ortiz}, M.~C., {et~al.}
  2017{\natexlab{b}}, MNRAS, 464, 3040

\end{thebibliography}

\end{document}